\documentclass[journal,twocolumn]{IEEEtran}
\usepackage{color}
\usepackage{graphicx}
\usepackage{epstopdf}
\usepackage{amsmath}
\usepackage{amssymb}
\usepackage{hyperref}
\usepackage[english]{babel}
\usepackage{cite}
\usepackage{rotfloat}
\usepackage{mathtools}
\usepackage[justification=centering,font=small,skip=0pt]{caption}

\usepackage{amsmath}
\usepackage{makecell}
\usepackage{algorithm,algorithmic}
\usepackage{multirow}
\usepackage{subfigure}
\usepackage{booktabs}
\usepackage{colortbl}
\usepackage{multirow}% http://ctan.org/pkg/multirow
\usepackage{hhline}% http://ctan.org/pkg/hhline

\definecolor{kugray5}{RGB}{224,224,224}

\usepackage[normalem]{ulem}
\newcommand\rsout{\bgroup\markoverwith
	{\textcolor{red}{\rule[0.5ex]{2pt}{0.8pt}}}\ULon}



\makeatletter
\newcommand{\ALOOP}[1]{\ALC@it\algorithmicloop\ #1%
	\begin{ALC@loop}}
	\newcommand{\ENDALOOP}{\end{ALC@loop}\ALC@it\algorithmicendloop}

\makeatother

\usepackage{etoolbox}
\let\mybibitem\bibitem
\renewcommand{\bibitem}[1]{%
	\ifstrequal{#1}{nature}
	{\color{blue}\mybibitem{#1}}
	{\color{black}\mybibitem{#1}}%
}

\graphicspath{ {Figures/} }

\newtheorem{remark}{Remark}

\DeclareCaptionLabelSeparator{periodspace}{.\quad}

\renewcommand{\figurename}{Fig.}
\addto\captionsenglish{\renewcommand{\figurename}{Fig.}}
\newcommand\nbthis{\addtocounter{equation}{1}\tag{\theequation}}

\newcommand{\norm}[1]{\left\lVert#1\right\rVert} % ||.||
\newcommand{\eq}[1]{\begin{align*}#1\end{align*}} % equation
\newcommand{\eqn}[1]{\begin{align}#1\end{align}} % equation
\newcommand{\nt}[1]{\left(#1\right)} % ()
\newcommand{\abs}[1]{\left|#1\right|} % ||
\newcommand{\re}[1]{\mathfrak{R}{\left(#1\right)}}
\newcommand{\im}[1]{\mathfrak{I}{\left(#1\right)}}

\newcommand{\mean}[1]{\mathbb{E} \left\{#1\right\}}

\newcommand{\mQ}{\textbf{\textit{Q}}}
\newcommand{\mR}{\textbf{\textit{R}}}
\newcommand{\mH}{\textbf{\textit{H}}}

\newcommand{\mW}{\textbf{\textit{W}}}
\newcommand{\mP}{\textbf{\textit{P}}}

\newcommand{\setR}{\mathbb{R}}
\newcommand{\setA}{\mathcal{A}}

\newcommand{\setAtld}{\tilde{\setA}^{N_t}}

\newcommand{\ve}{\textbf{\textit{e}}} 
\newcommand{\vs}{\textbf{\textit{s}}}
\newcommand{\vx}{\textbf{\textit{x}}}
\newcommand{\vy}{\textbf{\textit{y}}}

\newcommand{\vn}{\textbf{\textit{n}}}

\newcommand{\vz}{\textbf{\textit{z}}} 
\newcommand{\vh}{\textbf{\textit{h}}}

\newcommand{\vb}{\textbf{\textit{b}}}
\newcommand{\vw}{\textbf{\textit{w}}}
\newcommand{\va}{\textbf{\textit{a}}}
\newcommand{\vd}{\textbf{\textit{d}}}

\newcommand{\smt}{\sigma_t^2} % Power
\newcommand{\smn}{\sigma_n^2} % Power
\begin{document}
	\title{Application of Deep Learning to Sphere Decoding for Large MIMO Systems}
	\author{Nhan Thanh Nguyen, \IEEEmembership{Member, IEEE},
		Kyungchun Lee, \IEEEmembership{Senior Member, IEEE},
		and Huaiyu Dai, \IEEEmembership{Fellow, IEEE}

		\thanks{N. T. Nguyen was with the Department of Electrical and Information Engineering, Seoul National University of Science and Technology. He is currently with Centre for Wireless Communications, University of Oulu, P.O.Box 4500, FI-90014, Finland, (e-mail: nhan.nguyen@oulu.fi).}
		\thanks{K. Lee is with the Department of Electrical and Information Engineering and the Research Center for Electrical and Information Technology, Seoul National University of Science and Technology, 232 Gongneung-ro, Nowon-gu, Seoul, 01811, Republic of Korea (e-mail: kclee@seoultech.ac.kr).}
		\thanks{H. Dai is with the Department of Electrical and Computer Engineering, North Carolina State University, NC, USA. (e-mail: Huaiyu\_Dai@ncsu.edu).}
	}
	
	\maketitle
	\vspace{-2cm}
	\begin{abstract}
		Although the sphere decoder (SD) is a powerful detector for multiple-input multiple-output (MIMO) systems, it has become computationally prohibitive in massive MIMO systems, where a large number of antennas are employed. To overcome this challenge, we propose fast deep learning (DL)-aided SD (FDL-SD) and fast DL-aided $K$-best SD (KSD, FDL-KSD) algorithms. Therein, the major application of DL is to generate a highly reliable initial candidate to accelerate the search in SD and KSD in conjunction with candidate/layer ordering and early rejection. Compared to existing DL-aided SD schemes, our proposed schemes are more advantageous in both offline training and online application phases. Specifically, unlike existing DL-aided SD schemes, they do not require performing the conventional SD in the training phase. For a $24 \times 24$ MIMO system with QPSK, the proposed FDL-SD achieves a complexity reduction of more than $90\%$ without any performance loss compared to conventional SD schemes. For a $32 \times 32$ MIMO system with QPSK, the proposed FDL-KSD only requires $K = 32$ to attain the performance of the conventional KSD with $K=256$, where $K$ is the number of survival paths in KSD. This implies a dramatic improvement in the performance--complexity tradeoff of the proposed FDL-KSD scheme.
	\end{abstract}
	
	\begin{IEEEkeywords}
		massive MIMO, deep learning, deep neural network, sphere decoding, $K$-best sphere decoding.
	\end{IEEEkeywords}
	\IEEEpeerreviewmaketitle
	
	\section{Introduction}
	
	In mobile communications, the total throughput can be significantly enhanced by simultaneously transmitting/receiving as many data streams as possible with a large number of transmit/receive antennas. Therefore, multiple-input multiple-output (MIMO) technology may dramatically improve a system's spectral and energy efficiency. In a practical uplink multiuser large MIMO system, the numbers of transmit and receive antennas can be comparable \cite{rusek2013scaling}. In this scenario, low-complexity sub-optimal detectors, such as linear zero forcing (ZF), minimum-mean-square-error (MMSE), and successive interference cancellation (SIC) receivers, cannot achieve full diversity \cite{vardhan2008low}. In contrast, the maximum likelihood (ML) detector performs optimally, but its computational cost increases exponentially with the number of transmit antennas, which is prohibitive in large MIMO systems. Therefore, low-complexity near-optimal detection is an important challenge in optimizing large MIMO systems \cite{rusek2013scaling, chockalingam2014large, mandloi2017low}.
	
	\subsection{Related Works}
	
	The sequential sphere decoder (SD) with reduced complexity and near-optimal performance with respect to (w.r.t.) the optimal ML detector has been optimized well for small- and moderate-size MIMO systems. Among its variants, the Schnorr--Euchner SD (SE-SD) \cite{schnorr1994lattice, agrell2002closest} has the same performance as the conventional Fincke--Pohst SD (FP-SD) \cite{fincke1985improved, hassibi2005sphere} with reduced complexity. However, its complexity remains very high in large MIMO systems \cite{nguyen2019qr}. To address the problems of sequential SD, $K$-best SD (KSD) \cite{guo2006algorithm} was proposed to achieve fixed and reduced complexity. However, this algorithm suffers performance degradation, and does not ensure complexity reduction at high signal-to-noise ratios (SNRs).
	
	The aforementioned challenges of the sequential SD and KSD make them infeasible for large MIMO systems. However, the increasing application of deep learning (DL) in wireless communication creates room for further optimization of SD schemes. Particularly, the initial works on DL-aided SD in \cite{askri2019dnn} and \cite{mohammadkarimi2018deep} attempt to improve SD by employing a deep neural network (DNN) to learn the initial radius. Whereas a single radius (SR) is used in \cite{askri2019dnn}, multiple radii (MR) are employed in \cite{mohammadkarimi2018deep}. In this study, to distinguish them, we refer to the former as the SR-DL-SD scheme and to the latter as the MR-DL-SD scheme. Furthermore, as an improvement of \cite{askri2019dnn} and \cite{mohammadkarimi2018deep}, Weon et al. \cite{weon2020learning} propose a learning-aided deep path-prediction scheme for sphere decoding (DPP-SD) in large MIMO systems. Specifically, the minimum radius for each sub-tree is learned by a DNN, resulting in more significant complexity reduction w.r.t. the prior SR-DL-SD and MR-DL-SD schemes. {The application of DL to symbol detection in MIMO systems is not limited to the aforementioned DL-aided SD schemes. For example, various DL models have been proposed to directly estimate the transmitted signal vector \cite{nguyen2019deep,samuel2019learning,he2018model,khani2019adaptive,gao2018sparsely,samuel2017deep,wei2020learned, takabe2019trainable, li2020deep, sun2019learning, he2020model}. In general, these schemes have been shown to perform better than traditional linear detectors, such as ZF and MMSE, with low complexity. We further discuss these schemes in Section III-A.}
	
	In all three DL-aided SD schemes mentioned above, the common idea is to predict radii for the sequential SD. This approach has some limitations in the offline learning phase, as well as the online application. First, in the DNN training phase in \cite{askri2019dnn, mohammadkarimi2018deep,weon2020learning}, conventional SD needs to be performed first to generate training labels, i.e., the radius. Consequently, time and computational complexity requirements are high to train these DNNs. Although the training phase can be performed offline, these time and resource requirements make such schemes less efficient. Second, although the radius plays an important role in the search efficiency of conventional FP-SD, it becomes less significant in SE-SD \cite{hassibi2005sphere}. Therefore, using the predicted radius becomes less efficient in SE-SD, especially for high SNRs, for which a relatively reliable radius can be computed using the conventional formula \cite{hassibi2005sphere}. Moreover, in the KSD, the breadth-first search does not require a radius, which implies that the learning objectives in \cite{askri2019dnn, mohammadkarimi2018deep,weon2020learning} are inapplicable to the KSD scheme.
	
	\subsection{Contributions}
	
	In this study, we propose the fast DL-aided SD (FDL-SD) and fast DL-aided KSD (FDL-KSD) algorithms, which can overcome the limitations of the existing DL-aided SD schemes via a novel application of DL to SD. Specifically, we use a DNN to generate a highly reliable initial candidate for the search in SD, rather than generating the radius as in the existing DL-aided SD schemes \cite{askri2019dnn, mohammadkarimi2018deep,weon2020learning}. Furthermore, the output of the DNN facilitates a candidate/layer-ordering scheme and an early rejection scheme to significantly reduce the complexity. We note that the sequential SD and KSD have their own advantages and disadvantages. Specifically, the former guarantees near-optimal performance at the price of high computational complexity. By contrast, the KSD with reduced complexity can have performance degradation. In particular, the performance--complexity tradeoff of these schemes significantly depends on design parameters, which are the radius in the sequential SD and the number of surviving paths, i.e., $K$, in KSD. {In this work, we propose leveraging the fast-convergence sparsely connected detection network (FS-Net), a DNN architecture that was introduced in \cite{nguyen2019deep}, to further optimize their performance--complexity tradeoff}, and at the same time, to mitigate the dependence on the radius and $K$. Our specific contributions can be summarized as follows:
	
	\begin{itemize}
		
		\item For the application of DL to the SD scheme, rather than predicting the radius, we propose applying the FS-Net to generate a reliable solution to accelerate the SD scheme. Unlike other architecture that uses DNNs for learning the radius \cite{askri2019dnn, mohammadkarimi2018deep, weon2020learning}, the FS-Net can be trained easily without performing conventional SD; this considerably reduces the time and computational resources required for the training phase.
		
		\item We propose the FDL-SD scheme, which achieves significant complexity reduction while fully preserving the performance of the conventional SD. Specifically, we exploit the output of the FS-Net to facilitate the search in SD based on the following ideas:
		
		\textit{(i)} First, the output of the FS-Net, which is the approximate of the transmitted signal vector, is employed to determine the search order in the SD scheme. In particular, the candidates are ordered such that those closer to the FS-Net's output are tested first. This approach enhances the chance that the optimal solution is found early and accelerates the shrinking of the sphere, resulting in complexity reduction of the proposed FDL-SD scheme. 
		
		\textit{(ii)} Second, we propose a layer-ordering scheme. Specifically, we found that the sequential tree-like search in SD can be considered as the process of exploring and correcting incorrectly detected symbols. This implies that the errors at the lower layers can be explored and corrected sooner. Motivated by this, we propose ordering the layers of candidates so that errors are more likely to occur at low layers. This order is determined based on the FS-Net's output.
		
		\item In the proposed FDL-KSD scheme, the FS-Net's output is also leveraged to optimize the search process of the KSD. In this scheme, we employ the cost metric of the FS-Net-based solution as a threshold to reject unpromising candidates early. Furthermore, the layer ordering in \textit{(ii)} is used to reduce the chance that the optimal solution is rejected early. This results in not only performance improvement, but also complexity reduction w.r.t. the conventional KSD.
		
		\item Our extensive simulation results show that the FDL-SD scheme achieves a remarkable complexity reduction without any performance loss w.r.t. the conventional SD. In particular, the complexity reduction attained by our proposed FDL-SD scheme is significantly greater than those acquired by the existing DL-aided SD schemes. Furthermore, the proposed FDL-KSD scheme exhibits a considerable improvement in the performance--complexity tradeoff w.r.t. the conventional KSD.
	\end{itemize}
	
	We note that the aforementioned applications \textit{(i)} and \textit{(ii)} of DL to SD/KSD are not limited by the use of the FS-Net. They can also operate with an initial solution obtained by other linear or DL-based detectors. However, we found that FS-Net can be highly efficient for generating an initial solution for the proposed FDL-SD/KSD schemes, thus yielding significant complexity reduction. Specifically, the more reliable the initial solution, the greater the complexity reduction gain that can be achieved by the FDL-SD/KSD. However, it is worth noting that the computational complexity required to generate the initial solution must be included in the overall complexity of the FDL-SD/KSD. Therefore, the initial solution should be generated by a detector with a superior performance--complexity tradeoff, such as FS-Net. We further discuss this issue in Section III-A.
	
	In general, the integration of the FS-Net with the SD/KSD in this work and that with the tabu search (TS) scheme in \cite{nguyen2019deep} are common in enabling a favorable initialization of the search. However, they are based on different motivations and ideas, and also have different efficiencies. Specifically, in the TS, the search starts from a candidate and moves over its neighbors successively to find a near-optimal solution. Because a very large number of moves are required to ensure that a near-optimal solution is attained in massive MIMO systems, it is necessary to start from a reliable point and terminate the search early. In \cite{nguyen2019deep}, we proposed using the FS-Net output as the initial point for moving among neighbors, and based on its quality, the search can be terminated early to reduce the complexity. In this sense, the search can end before the optimal solution is reached, causing performance loss for DL-TS. In contrast, the proposed FDL-SD/KSD schemes find exactly the same solution as their conventional counterparts, but much faster. To this end, candidate/layer ordering is proposed to accelerate the shrinking of the hypersphere to reach the solution as soon as possible; the search process is not terminated early. As a result, the FDL-SD fully preserves the performance, whereas the FDL-KSD provides improved performance w.r.t. the conventional SD/KSD.
	
	\textit{Paper structure}: The rest of the paper is organized as follows: Section II presents the system model. In Section III, the existing DNNs for MIMO detection are reviewed, and the FS-Net's architecture and operation are described. The proposed FDL-SD and FDL-KSD schemes are presented in Sections IV and V, respectively. In Section VI, the simulation results and numerical discussions are presented. Finally, conclusions are drawn in Section VII.
	
	\textit{Notations}: Throughout this paper, scalars, vectors, and matrices are denoted by lowercase, bold-face lowercase, and bold-face uppercase letters, respectively. The $i$th element of a vector $\va$ is denoted by $a_i$, and the $(i,j)$th element of a matrix $\textbf{\textit{A}}$ is denoted by $a_{i,j}$. $(\cdot)^T$ denotes the transpose of a matrix. Furthermore, $\abs{\cdot}$ and $\norm{\cdot}$ represent the absolute value of a scalar and the Frobenius norm of a matrix, respectively; $\sim$ means $\textit{distributed as}$.
	
	\section{{System Model and MIMO Detection}}
	\subsection{{System Model}}
	Consider the uplink of a multiuser MIMO system, where the base station is equipped with $N_r$ receive antennas, and the total number of transmit antennas among all users is $N_t$, $N_r \geq N_t$. The received signal vector $\tilde{\vy}$ is given by
	\eqn{
		\label{complex SM}
		\tilde{\vy} = \tilde{\mH} \tilde{\vs} + \tilde{\vn},
	}
	where $\tilde{\vs} = [\tilde{s}_1, \tilde{s}_2, \ldots, \tilde{s}_{N_t}]$ is the vector of transmitted symbols with $\mean{\abs{\tilde{s}_i}^2} = \smt$. The transmitted symbols $\tilde{s}_i, i = 1, 2, \ldots, N_t,$ are drawn independently from a complex constellation $\tilde{\setA}$ of $\tilde{Q}$ points. The set of all possible transmitted vectors forms an $N_t$-dimensional complex constellation $\setAtld$ consisting of $\tilde{Q}^{N_t}$ vectors, i.e., $\tilde{\vs} \in \setAtld$. In \eqref{complex SM}, $\tilde{\vn}$ is a vector of independent and identically distributed (i.i.d.) additive white Gaussian noise (AWGN) samples, i.e., $\tilde{n}_i \sim \mathcal{CN}(0,\smn)$. Furthermore, $\tilde{\mH}$ denotes an $N_r \times N_t$ channel matrix consisting of entries $\tilde{h}_{i,j}$, where $\tilde{h}_{i,j}$ represents the complex channel gain between the $j$th transmit antenna and $i$th receive antenna. 
	
	Let $\vs, \vy, \vn,$ and $\mH$ denote the $\nt{M \times 1}$-equivalent real transmitted signal vector, $\nt{N \times 1}$-equivalent real received signal, $\nt{N \times 1}$-equivalent real AWGN noise signal vectors, and $(N \times M)$-equivalent real channel matrix, respectively, with $M = 2N_t, N = 2N_r$, where 
	\eq{
		\vs = 
		\begin{bmatrix}
			\re{\tilde{\vs}}\\
			\im {\tilde{\vs}}
		\end{bmatrix}, \hspace{0.1cm}
		\vy = 
		\begin{bmatrix}
			\re{\tilde{\vy}}\\
			\im {\tilde{\vy}}
		\end{bmatrix}, \hspace{0.1cm}
		\vn = 
		\begin{bmatrix}
			\re{\tilde{\vn}}\\
			\im {\tilde{\vn}}
		\end{bmatrix},
	}
	and
	\begin{align*}
		\mH = 
		\begin{bmatrix}
			\re {\tilde{\mH}}  &-\im {\tilde{\mH}}\\
			\im {\tilde{\mH}}  &\re {\tilde{\mH}}
		\end{bmatrix}.
	\end{align*}
	Here, $\re {\cdot}$ and $\im {\cdot}$ denote the real and imaginary parts of a complex vector or matrix, respectively. In practical communication systems, high-order QAM such as 16-QAM and 64-QAM is more widely employed than high-order PSK modulation schemes. Therefore, we assume that 16-QAM and 64-QAM schemes are employed for high-order modulations, whereas QPSK is considered for low-order modulation. Then, the complex signal model \eqref{complex SM} can be converted to an equivalent real-signal model
	\eqn{
		\vy = \mH \vs + \vn. \label{real SM}
	}
	The set of all possible real-valued transmitted vectors forms an $M$-dimensional constellation $\setA^M$ consisting of $Q^M$ vectors, i.e., $\vs \in \setA^M$, where $\setA$ is the real-valued symbol set. In this study, we use the equivalent real-valued signal model in \eqref{real SM} because it can be employed readily for both SD algorithms and DNNs.
	
	\subsection{{Detection in MIMO Systems}}
	\subsubsection{{Conventional optimal solution}}
	The ML solution can be written as
	\eqn {
		\hat{\vs}_{ML} = \arg \min_{\vs \in \setA^{M}} \norm {\vy - \mH \vs}^2. \label{ML_solution}
	}
	The computational complexity of ML detection in \eqref{ML_solution} is exponential with $M$ \cite{nguyen2019qr, nguyen2019deep}, which results in extremely high complexity for large MIMO systems, where $M$ is very large. 
	
	\subsubsection{{DNN-based solution}}
	
	Consider a DNN of $L$ layers with input vector $\vx^{[1]}$, including the information contained in $\vy$ and $\mH$, and the output vector $\hat{\vs}^{[L]}$. Let $\hat{\vs} = \mathcal{Q} \left( \hat{\vs}^{[L]} \right)$,
	where $\mathcal{Q} (\cdot)$ is the element-wise quantization operator that quantizes $ \hat{s}^{[L]}_m \in \setR$ to $\hat{s}_m \in \setA, m=1, \ldots,M$. The DNN can be trained so that $\hat{\vs}$ is an approximate of the transmitted signal vector $\vs$. In the DNN, serial nonlinear transformations are performed to map $\vx^{[1]}$ to $\hat{\vs}^{[L]}$ as follows:
	\begin{align*}
	\hat{\vs}^{[L]} = f^{[L]} \left( \ldots \left(f^{[1]} \left(\vx^{[1]}; \mP^{[1]} \right); \ldots \right); \mP^{[L]} \right), \nbthis \label{DNN_1}
	\end{align*}
	where
	\begin{align*}
	f^{[l]} \left(\vx^{[l]}; \mP^{[l]} \right) = \sigma^{[l]} \left( \mW^{[l]} \vx^{[l]} + \vb^{[l]} \right) \nbthis \label{f^{[l]}}
	\end{align*}
	represents the nonlinear transformation in the $l$th layer with the input vector $\vx^{[l]}$, activation function $\sigma^{[l]}$, and $\mP^{[l]} = \left\{\mW^{[l]}, \vb^{[l]}\right\}$ consisting of the weighting matrix $\mW^{[l]}$ and bias vector $\vb^{[l]}$ {whose size depends on the structure of the input vector $\vx^{[l]}$}.
	
	The computational complexity of a DNN depends on its depth and the number of neurons in each layer, which are both determined by the size of the input vector. {These are usually  optimized and selected by simulations; however, in general, a larger input vector requires a deeper DNN and/or more neurons in each layer.} In a DNN for large MIMO detection, the input is a high-dimensional vector because it contains the information of a large-size channel matrix and the received signal vector. As a result, large-size weight matrices and bias vectors, i.e., $\mW^{[l]}$ and $\vb^{[l]}$, are required in \eqref{f^{[l]}}. Furthermore, in large MIMO systems, many hidden layers and neurons are required for the DNN to extract meaningful features and patterns from the large amount of input data and provide high accuracy. Therefore, the computational complexity of the detection network typically becomes very high in large MIMO systems. 
	
	\section{DNNs for MIMO Detection and the FS-Net}

	In this section, we review the state-of-the-art DNN architectures for MIMO detection and explain why FS-Net is chosen for incorporation with the proposed FDL-SD and FDL-KSD schemes. Then, the network architecture and operation of the FS-Net are briefly introduced.
	
	\subsection{DNNs for MIMO Detection}
	
	A number of deep neural networks (DNNs) have been designed for symbol detection in large MIMO systems \cite{samuel2019learning, nguyen2019deep, he2018model, khani2019adaptive, gao2018sparsely, samuel2017deep, wei2020learned, takabe2019trainable, li2020deep, sun2019learning}. Specifically, Samuel et al. in \cite{samuel2019learning} and \cite{samuel2017deep} introduced the first DNN-based detector, called the detection network (DetNet). However, the DetNet performs poorly for large MIMO systems with $M \approx N$; it also has a complicated network architecture with high computational complexity. To overcome these challenges, the sparsely connected detection network (ScNet) \cite{gao2018sparsely} and FS-Net \cite{nguyen2019deep} {have been} proposed. They simplify the network architecture and improve the loss function of the DetNet, which leads to significant performance improvement and complexity reduction. Furthermore, in \cite{wei2020learned}, a learned conjugate gradient descent network (LcgNet) is proposed. DetNet, ScNet, FS-Net, and LcgNet are similar in the sense that they are obtained by unfolding the iterative gradient descent method. A trainable projected gradient-detector (TPG-detector) was proposed in \cite{takabe2019trainable} to improve the convergence of the projected gradient method. Recently, DL-aided detectors based on iterative search, including the DL-based likelihood ascent search (DPLAS) and learning to learn the iterative search algorithm (LISA), were proposed in \cite{li2020deep} and \cite{sun2019learning}, respectively. By unfolding the orthogonal approximate message passing (OAMP) algorithm \cite{ma2017orthogonal}, He et al. introduced the OAMP-Net \cite{he2018model,he2020model} for symbol detection in both i.i.d. Gaussian and small-size correlated channels. Furthermore, Khani et al. in \cite{khani2019adaptive} focus on realistic channels and propose the MMNet, which significantly outperforms the OAMP-Net with the same or lower computational complexity. 
	
	The main application of DL to SD in this work is to generate a highly reliable candidate $\hat{\vs}$ that is an approximate of the transmitted signal vector $\vs$. This can be achieved by any of the aforementioned DNNs, i.e., DetNet, ScNet, FS-Net, OAMP-Net, MMNet, and LcgNet. In this work, we chose FS-Net because of its low complexity and reliable BER performance. Specifically, among the discussed DNNs, the iterative schemes, i.e., the OAMP-Net and MMNet, require the highest computational complexity because pseudo-matrix inversion is performed in each layer to conduct the linear MMSE estimation \cite{he2018model}, and/or to compute the standard deviation of the Gaussian noise on the denoiser inputs in each layer \cite{khani2019adaptive}. Meanwhile, DetNet employs a dense connection architecture with high-dimensional input vectors in every layer \cite{samuel2019learning,samuel2017deep}, which causes an extremely high computational load. In contrast, FS-Net has a superior performance--complexity tradeoff. To achieve this, its network architecture is optimized to become very sparse, whereas its loss function is optimized for fast convergence. We also note that the FS-Net can output a reliable solution with only element-wise matrix multiplications, and no matrix inversion is required. Furthermore, the simulation results in \cite{nguyen2019deep} show that FS-Net achieves better performance with lower complexity compared to DetNet, ScNet, and Twin-DNN. Therefore, it is chosen for incorporation with the SD schemes in this work.
	
	\subsection{Network Architecture and Operation of the FS-Net}

	\begin{figure*}[t]
		\centering
		\subfigure[FS-Net architecture]
		{
			\includegraphics[scale=0.98]{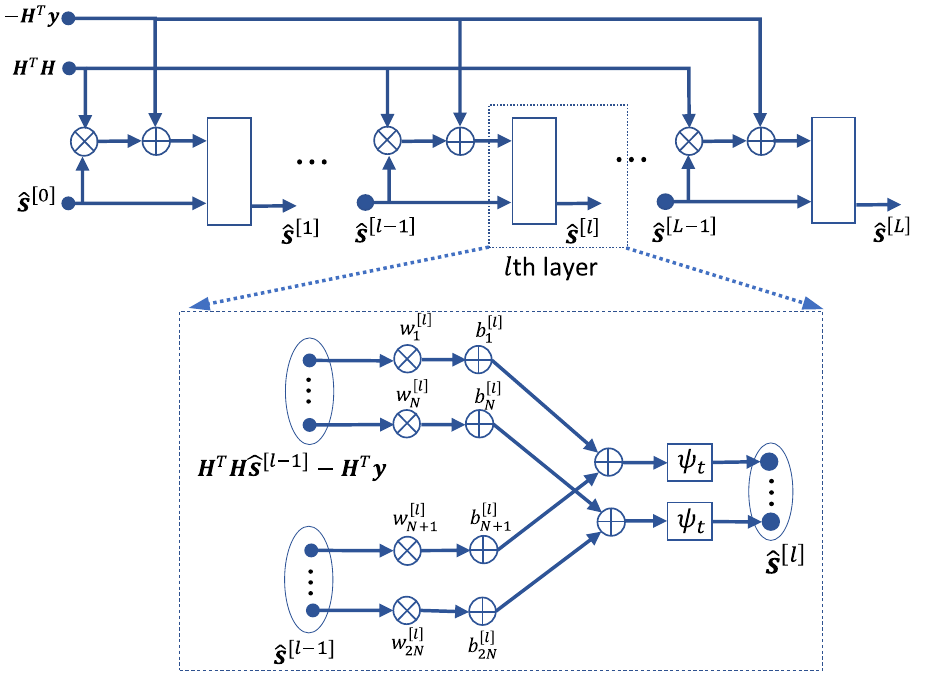}
			\label{fig_dscnet}
		}
		\subfigure[$\psi_t(x)$ with $t = 0.5$]
		{
			\includegraphics[scale=0.55]{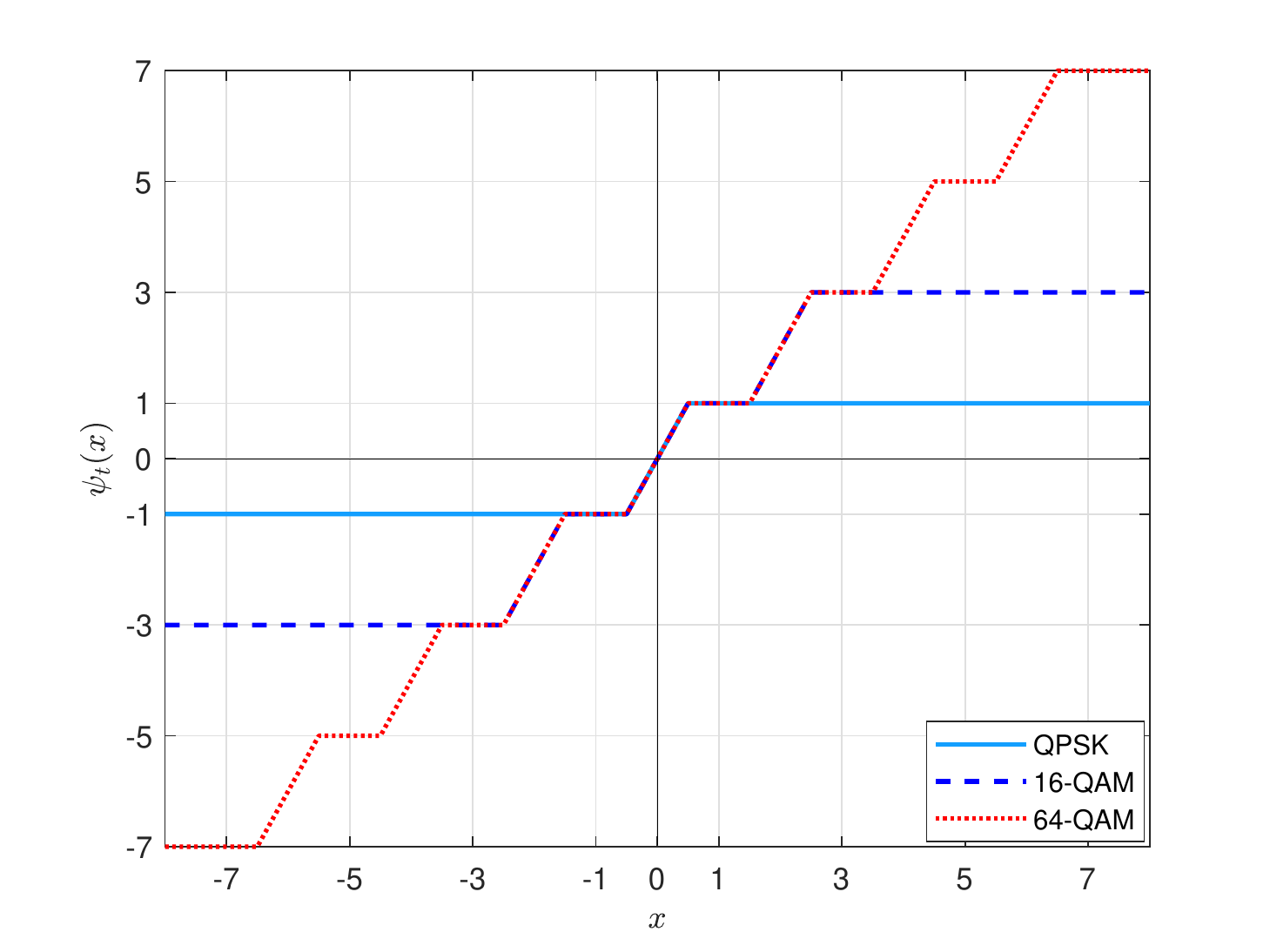}
			\label{fig_phi}
		}
		\caption{Illustration of the FS-Net architecture and $\psi_t(x)$ in FS-Net for QPSK, 16-QAM, and 64-QAM.}
	\end{figure*}
	
	\begin{algorithm}[t]
		\caption{FS-Net scheme for MIMO detection}
		\label{al_fsnet}
		\begin{algorithmic}[1]
			\REQUIRE $\mH, \vy$.
			\ENSURE $\hat{\vs}$.
			\STATE Compute $\mH^T \mH$ and $\mH^T \vy$.
			\STATE {$\hat{\vs}^{[0]} = \textbf{0}$}
			\FOR {$l = 1 \rightarrow L$}
			\STATE $\vz^{[l]} = \mH^T \mH \hat{\vs}^{[l-1]} - \mH^T \vy$
			\STATE {$\hat{\vs}^{[l]} = \psi_t \left(\vw_1^{[l]} \odot \hat{\vs}^{[l-1]} + \vb_1^{[l]}\right) + \psi_t \left(\vw_2^{[l]} \odot \vz^{[l]} + \vb_2^{[l]}\right)$}
			\ENDFOR 
			\STATE {$\hat{\vs} = \mathcal{Q} \left( \hat{\vs}^{[L]} \right)$}
		\end{algorithmic}
	\end{algorithm}

	In the FS-Net, $\hat{\vs}$ is updated over $L$ layers of the DNN by mimicking a projected gradient descent-like ML optimization as follows \cite{samuel2019learning, samuel2017deep}:
	\begin{align*}
	\hat{\vs}^{[l+1]} &= \Pi \left[ \vs - \delta^{[l]} \frac{\partial \norm {\vy - \mH \vs}^2}{\partial \vs} \right]_{\vs = \hat{\vs}^{[l]}} \\
	&= \Pi \left[ \hat{\vs}^{[l]} + \delta^{[l]} (\mH^T \mH \hat{\vs}^{[l]} - \mH^T \vy) \right], \nbthis \label{fsnet_0}
	\end{align*}
	where $\Pi[\cdot]$ denotes a nonlinear projection operator and $\delta^{[l]}$ is a step size.
	
	The network architecture of the FS-Net is illustrated in Fig. \ref{fig_dscnet}. The operation of the FS-Net is summarized in Algorithm \ref{al_fsnet}, where in step 5, $\{\vw_1^{[l]}, \vb_1^{[l]}, \vw_2^{[l]}, \vb_2^{[l]} \} \in \setR^{M \times 1}$ represents the weights and biases of the FS-Net in the $l$th layer, and $\odot$ denotes the element-wise multiplication of two vectors. Furthermore, $\psi_t(\cdot)$ is used to guarantee that the amplitudes of the elements of $\hat{\vs}^{[l]}$ are within the range of the corresponding modulation size, such as $[-1,1]$ for QPSK, $[-3,3]$ for 16-QAM, and $[-7,7]$ for 64-QAM, as illustrated in Fig. \ref{fig_phi}. In particular,
	\begin{align*}
	\psi_t(x) = -q + \frac{1}{\abs{t}} \sum_{i \in \boldsymbol{\Omega}}[\sigma(x + i + t) - \sigma(x + i- t)]
	\end{align*}
	with $q=1$, $\boldsymbol{\Omega} = \{0\}$ for QPSK, $q=3$, $\boldsymbol{\Omega} = \{-2,0,2\}$ for 16-QAM, $q=7$, $\boldsymbol{\Omega} = \{-6,-4,-2,0,2,4,6\}$ for 64-QAM, and $\sigma(\cdot)$ is the rectified linear unit (ReLU) activation function. The final detected symbol vector is given as $\hat{\vs} = \mathcal{Q} \left( \hat{\vs}^{[L]} \right)$ in step 7 of Algorithm \ref{al_fsnet}, where $\mathcal{Q} (\cdot)$ represents an element-wise quantization function. During the training of the FS-Net, the following loss function is used \cite{nguyen2019deep}: 
	\begin{align*}
		\mathcal{L} (\vs, \hat{\vs}) = \sum_{l=1}^{L} \log (l) \left[ \norm{\vs - \hat{\vs}^{[l]}}^2 + \xi r (\hat{\vs}^{[l]}, \vs) \right], \nbthis \label{loss_dscnet}
	\end{align*}
	where $\hat{\vs}^{[l]}$ and $\vs$ are the output of the $l$th layer and the desired transmitted signal vector, respectively, and $r (\hat{\vs}^{[l]}, \vs) = 1 - (|\vs^T \hat{\vs}^{[l]}|) / (\norm{\vs}  \lVert \hat{\vs}^{[l]}\rVert) $.
	
	In the online application phase, the overall complexity of the FS-Net can be given as
	\begin{align*}
	\mathcal{C}_{\text{FS-Net}} = M(2N-1) + M^2(2M-1) + L(2N^2 + 5N ), \nbthis \label{comp_fsnet}
	\end{align*}
	where the first, second, and last terms are the total numbers of additions and multiplications required for the computations of $\mH^T \vy$, $\mH^T \mH$, and the processing in the $L$ layers of the FS-Net, respectively.

	\section{Proposed FDL-SD Scheme}
	
	\begin{figure*}[t]
		\centering
		\includegraphics[scale=0.55]{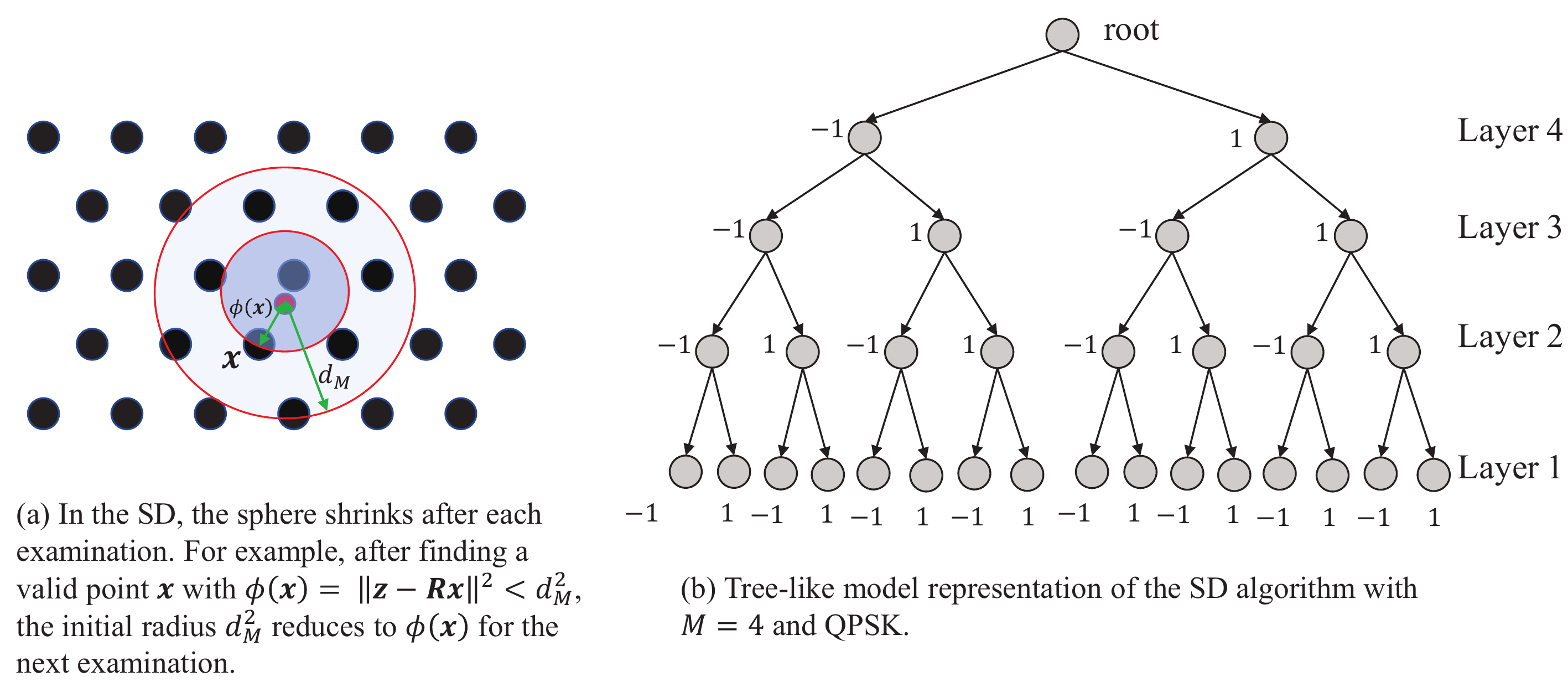}
		\caption{Illustration of the SD scheme with the lattice and tree-like model.}
		\label{fig_tree_model}
	\end{figure*}

	We first briefly review the common ideas of SD based on the description in \cite{hassibi2005sphere}. Some notations in \cite{hassibi2005sphere} are also adopted for ease of notation. Similar to the ML detection, SD attempts to find the optimal lattice point that is closest to $\vy$, but its search is limited to the points inside a sphere of radius $d$, i.e.,
	\eqn {
		\hat{\vs}_{SD} = \arg \min_{\vx \in \mathcal{S} \subset \setA^{M}} \norm {\vy - \mH \vx}^2, \label{SD_solution}
	}
	where $\mathcal{S}$ is a hypersphere specified by the center $\vy$ and radius $d$, {and $\vx$ represents the hypothesized transmitted signal vector that lies inside $\mathcal{S}$}. Each time a valid lattice point, i.e., a point lying inside $\mathcal{S}$, is found, the search is further restricted, or equivalently, the sphere shrinks, by decreasing the radius, as illustrated in Fig. \ref{fig_tree_model}(a). In this way, when there is only one point in the sphere, the point becomes the final solution $\hat{\vs}_{SD}$. The ingenuity of SD is the identification of the lattice points that lie inside the sphere, which is discussed below.
	
	A lattice point $\mH \vx$ lies inside a sphere of radius $d$ if and only if $\vx$ fulfills condition $(\mathcal{C}): \norm{\vy - \mH \vx}^2 \leq d^2$. In SD, the QR decomposition of the channel matrix is useful in breaking $(\mathcal{C})$ into the necessary conditions for each element of $\vx$. Let $\mH = \mQ
	\begin{bmatrix}
	\mR \\
	\mathbf{0}_{(N-M) \times M}
	\end{bmatrix}$,
	where $\mQ = \left[ \mQ_1 \hspace{0.2cm} \mQ_2 \right] $ is an $N \times N$ unitary matrix, having the first $M$ and last $N-M$ orthonormal columns in $\mQ_1$ and $\mQ_2$, respectively. $\mR$ is an $M \times M$ upper triangular matrix, and $\mathbf{0}_{(N-M) \times M}$ represents a matrix of size $(N-M) \times M$ containing all zeros. Applying QR decomposition, $(\mathcal{C})$ can be rewritten as $\norm{\vz - \mR \vx}^2 \leq d_M^2$, where $\vz = \mQ_1^T \vy$ and $d_M^2 = d^2 - \norm{\mQ_2^T \vy}^2$. Owing to the upper-triangular structure of $\mR$, we have
	\begin{align*}
	(\mathcal{C}): \sum_{m=1}^{M} \left(z_m - \sum_{i=m}^{M} r_{m,i} x_i \right) ^2 \leq d_M^2.
	\end{align*}
	Consequently, the necessary conditions for the elements of $\vx$ to fulfill $(\mathcal{C})$ can be expressed as
	\begin{align*}
	LB_m \leq x_m \leq UB_m, m=1,2,\ldots,M, \nbthis \label{cond_cand}
	\end{align*}
	where $x_m$ represents the $m$th element of $\vx$. In \eqref{cond_cand}, $LB_m$ and $UB_m$ respectively denote the lower and upper bounds of $x_m$. Without loss of generality, we assume that the entries on the main diagonal of $\mR$ are positive, i.e., $r_{m,m} > 0, m=1,\ldots,M$. Then, $LB_m$ and $UB_m$ can be given as
	\begin{align*}
	LB_m = \left\lceil \frac{z_{m|m+1} - d_{m}}{r_{m,m}} \right\rceil, \hspace{0.2cm}
	UB_m = \left\lfloor \frac{z_{m|m+1} + d_{m}}{r_{m,m}} \right\rfloor, \nbthis \label{bound_m} 
	\end{align*}
	where $\lceil \cdot \rceil$ and $\lfloor \cdot \rfloor$ round a value to its nearest larger and smaller symbols in alphabet $\mathcal{A}$, respectively, and
	\begin{align*}
	z_{m|m+1} = 
	\begin{cases}
	z_M, & m=M \\
	z_m - \sum_{i=m+1}^{M} r_{m,i} x_i, & m<M
	\end{cases}
	\nbthis \label{z_m}
	\end{align*}
	means adjusting $z_m$ based on the chosen symbols of $\vx$, i.e., $\{x_{m+1}, x_{m+2}, \ldots, x_M\}$. Furthermore, $d_m$ in \eqref{bound_m} is given by
	\begin{align*}
	d_m^2 =
	\begin{cases}
	d_M^2, & m=M \\
	d_{m+1}^2 - \sigma_{m+1}^2, & m<M
	\end{cases},\nbthis \label{d_m}
	\end{align*}
	where
	\begin{align*}
	\sigma_{m+1}^2 = \left(z_{m+1|m+2} - r_{m+1,m+1} x_{m+1}\right)^2. \nbthis \label{sigma_m}
	\end{align*}
	
	The tree-like model is useful to illustrate the candidate exploration and examination in SD schemes. It maps all possible candidates $\vx \in \setA^{M}$ to a tree with $M$ layers, each associated with an element of $\vx$. Layer $m$ of the tree has nodes representing possibilities for $x_m$. A candidate is examined by extending a path, starting from the root, over nodes in the layers. When the lowest layer is reached, a complete path represents a candidate $\vx$. As an example, a tree-like model for a MIMO system with $M=4$ and QPSK is illustrated in Fig. \ref{fig_tree_model}(b), which has four layers, corresponding to $M=4$ elements of a candidate, and $\abs{\mathcal{A}}^M = 16$ complete paths, representing 16 candidates for the solution, where $\abs{\mathcal{A}}=2$ for QPSK signals. Based on the tree-like model, in the sequential SD, the candidates are explored in the depth-first search strategy. Specifically, the algorithm explores the nodes associated with the symbols satisfying \eqref{cond_cand} from the highest to lowest layers. Once the lowest layer of a candidate $\vx$ is reached, a valid lattice point is found, and the radius is reduced to $\phi(\vx)$ $\left(< d_M^2\right)$, where $\phi(\vx) = \norm{\vz - \mR \vx}^2 = d_M^2 - d_1^2 + (z_1 - r_{1,1} x_1)^2$ is the ML metric of $\vx$ \cite{hassibi2005sphere}, as shown in Fig. \ref{fig_tree_model}(a). Based on this search procedure, we found that SD can be optimized by ordering the examined candidates and layers in conjunction with the output of the FS-Net, as presented in the following subsections.
	
	\begin{figure*}[t]
		\centering
		\includegraphics[scale=0.55]{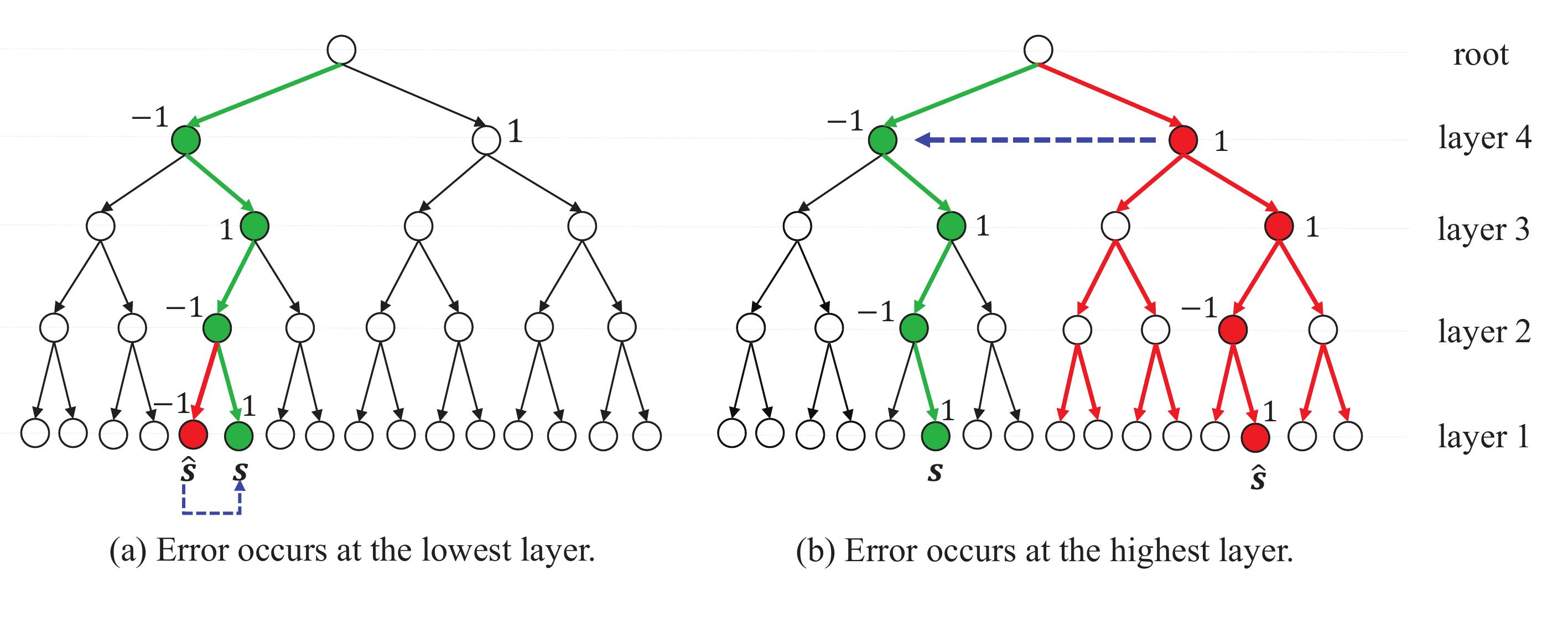}
		\caption{Illustration of the effect of erroneous layer on the search efficiency of SD. The dashed arrow presents the process of exploration and correction of the erroneous symbol.}
		\label{fig_tree_SESD}
	\end{figure*}

	\subsection{Candidate Ordering}
	
	The complexity of the SD scheme significantly depends on the number of lattice points that lie inside the sphere, or equivalently, the number of candidates that need to be examined. The sphere shrinks after a valid lattice point is found. Therefore, it is best to start the search by examining an optimal or near-optimal point. In the best case, if the algorithm starts with the optimal ML solution, i.e., $\vx = \hat{\vs}_{ML}$, the radius can decrease rapidly to $\phi(\hat{\vs}_{ML})$. As a result, no more lattice points lie inside the newly shrunken sphere, and the solution is concluded to be $\hat{\vs}_{SD} = \hat{\vs}_{ML}$. However, finding the optimal point requires high computational complexity, which is as challenging as performing the SD scheme itself. Furthermore, the simple linear ZF/MMSE or SIC detector cannot guarantee a highly reliable solution in practical multiuser large MIMO systems. Therefore, we propose using a DNN to find a reliable candidate for initializing the search in SD. For the reasons explained in Section III-A, the FS-Net is employed for this purpose.
	
	In the proposed FDL-SD scheme, the search starts by examination of $\hat{\vs}$ obtained in step 7 of Algorithm \ref{al_fsnet}. Furthermore, as $\hat{\vs}^{[L]}$ is the output of the FS-Net, it is natural to perform the search in an order such that a candidate closer to $\hat{\vs}^{[L]}$ is examined earlier. To this end, the symbols satisfying \eqref{cond_cand} in layer $m$ are ordered by increasing distance from $\hat{s}_m^{[L]}$, $m=1,2,\ldots,M$. Specifically, in layer $m$, the symbols are examined in the order
	\begin{align*}
	\mathcal{O}^{\text{FDL}}_m = \{ \hat{s}_m^{(1)}, \hat{s}_m^{(2)}, \ldots \}, \nbthis \label{order_FS}
	\end{align*}
	where $\hat{s}_m^{(i)} \in \mathcal{A} \text{ and } LB_m \leq \hat{s}_m^{(i)} \leq UB_m$, with $\hat{s}_m^{(i)}$ representing the $i$th-closest symbol to $\hat{s}_m^{[L]}$. Furthermore, $LB_m$ and $UB_m$ are given in  \eqref{bound_m}. By using $\{\mathcal{O}^{\text{FDL}}_1, \mathcal{O}^{\text{FDL}}_2, \ldots, \mathcal{O}^{\text{FDL}}_M\}$, the first candidate examined in the FDL-SD scheme is $\hat{\vs}$. It can be seen that in this scheme, the initial sphere is predetermined by the radius $\phi(\hat{\vs})$. However, in the case that $\hat{\vs}$ is unreliable, the sphere with radius $\phi(\hat{\vs})$ can be large and inefficient for the search. Therefore, the initial radius is set to $d^2 = \min \{\alpha N_r \sigma_n^2, \phi(\hat{\vs})\}$, where $\alpha N_r \sigma_n^2$ is the conventional radius \cite{hassibi2005sphere}.

	%It can be seen that in this scheme, the initial sphere is predetermined by the radius $\phi(\hat{\vs})$. Therefore, optimization of the initial radius is not required in the proposed FDL-SD scheme.
	
	It is worth noting that the proposed candidate ordering in the FDL-SD scheme is different from that in the SE-SD scheme. Specifically, the SE-SD scheme starts its search near the center of the sphere first, then moves outward to the surface of the sphere \cite{chan2002new}. Therefore, the order $\mathcal{O}^{\text{SE}}_m$ is employed in the SE-SD scheme\cite{chan2002new, vikalo2005sphere}, which is given by
	\begin{align*}
	\mathcal{O}^{\text{SE}}_m = \{ \bar{s}_m^{(1)}, \bar{s}_m^{(2)}, \ldots \}, \nbthis \label{order_SE}
	\end{align*}
	where $\bar{s}_m^{(i)} \in \mathcal{A}$ and $LB_m \leq \bar{s}_m^{(i)} \leq UB_m$. Here, $\bar{s}_m^{(i)}$ is the $i$th closest symbol to $\bar{s}_m = \frac{z_{m|m+1}}{r_{m,m}}$ and $LB_m$ and $UB_m$ are given in \eqref{bound_m}. Our simulation results show that the proposed FDL-SD scheme with order $\mathcal{O}^{\text{FDL}}_m$ results in considerable complexity reduction, which is much more significant than that provided by the SE-SD with order $\mathcal{O}^{\text{SE}}_m$.
	
	\subsection{Layer Ordering}

	In the SD scheme, once a candidate $\vx$ is examined, the next step is to search for a better solution in the shrunken sphere. This is equivalent to the process of correcting sub-optimally detected symbols in $\vx$, and the faster a wrong symbol is corrected, the earlier the optimal solution is found, which results in lower complexity of the SD scheme. Furthermore, knowledge of the positions of erroneous symbols can significantly affect the efficiency of error correction. Therefore, in this subsection, we focus on optimizing the layers of erroneous symbols, which will be referred to as \textit{erroneous layers} from now on.

	Similar to the conventional SD, in the proposed FDL-SD, the search repeatedly moves downward and upward over layers to explore candidates. Given that $\hat{\vs}$ is examined first and the order $\mathcal{O}^{\text{FDL}}_m$ in \eqref{order_FS} is used in each layer, the candidates are examined in the following order:
	\begin{align*}
	&{\begin{bmatrix} \hat{s}_M^{(1)} \\ \vdots \\ \hat{s}_m^{(1)} \\ \vdots \\ \hat{s}_1^{(1)}  \end{bmatrix}}
	\rightarrow \cdots 
	{\begin{bmatrix} \hat{s}_M^{(1)} \\ \vdots \\ \hat{s}_m^{(1)}  \\ \vdots \\ \hat{s}_1^{\left(\abs{\mathcal{O}^{\text{FDL}}_1}\right)} \end{bmatrix}} \rightarrow \ldots
	\rightarrow 
	{\begin{bmatrix} \hat{s}_M^{(1)} \\ \vdots \\ \hat{s}_m^{(2)} \\ \vdots  \end{bmatrix}}\\
	&\quad	\rightarrow \cdots
	{\begin{bmatrix} \hat{s}_M^{(1)} \\ \vdots \\ \hat{s}_m^{\left(\abs{\mathcal{O}^{\text{FDL}}_m}\right)} \\ \vdots  \end{bmatrix}}\rightarrow \cdots
	{\begin{bmatrix} \hat{s}_M^{(2)} \\ \vdots\end{bmatrix}}
	\rightarrow \cdots
	{\begin{bmatrix} \hat{s}_M^{\left(\abs{\mathcal{O}^{\text{FDL}}_M}\right)} \\ \vdots \end{bmatrix}}, \nbthis \label{layer_M}
	\end{align*}
	where the layers of candidates are reversed to reflect the tree-model-based search strategy in SD, and $\abs{\mathcal{O}^{\text{FDL}}_m}$ denotes the cardinality of $\mathcal{O}^{\text{FDL}}_m$. It is observed that, starting from $\hat{\vs} = \left[ \hat{s}_1^{(1)}, \ldots, \hat{s}_M^{(1)} \right]^T$, all the candidate symbols for layer $1$ in $\mathcal{O}^{\text{FDL}}_1$ are examined, and it moves upward to the higher layers and performs the same examination process, as shown in \eqref{layer_M}. From the candidate examination order above, we have a note in the following remark.
	\begin{remark}
		If an erroneous layer is $\tilde{m}$, i.e., $\hat{s}_{\tilde{m}} \neq s_{\tilde{m}}$, $\hat{s}_{\tilde{m}}$ can be corrected by replacing it with one of the other symbols in $\mathcal{O}^{\text{FDL}}_m$. In particular, the lower the erroneous layer is, the faster the corresponding erroneous symbol is corrected. An equivalent interpretation using the tree-like model is that, if there is an erroneous node, the closer to the leaf nodes it is, the faster it is corrected. For example, if $\tilde{m}=1$, the erroneous symbol $\hat{s}_1$ is represented by a leaf node, which can be corrected after examining at most $\abs{\mathcal{O}^{\text{FDL}}_1}-1$ other leaf nodes. In contrast, if $\tilde{m}=M$, the erroneous symbol $\hat{s}_M$ is represented by a node closest to the root, and it is only corrected after examining the entire sub-tree having $\hat{s}_M$ as the root. Consequently, much higher computational complexity is required than that required to examine only the leaf nodes, as in the case for $\tilde{m}=1$.
	\end{remark}
	
	According to Remark 1, the optimal solution can be found earlier if the errors exist at low layers. This can be illustrated in Fig. \ref{fig_tree_SESD} for a MIMO system with $M=4$ and QPSK signaling. We assume that the optimal solution is $\vs = [-1,1,-1,1]^T$ and that there is only a single erroneous symbol in $\hat{\vs}$. In Fig. \ref{fig_tree_SESD}(a), $\hat{\vs} = [-1,1,-1,-1]^T$, and the error occurs at the lowest layer, i.e., $\tilde{m}=1$, yielding $\hat{s}_1 \neq s_1$. It is observed that only one node is required to be examined to reach the optimal solution. In contrast, in Fig. \ref{fig_tree_SESD}(b), we assume $\hat{\vs} = [1,1,-1,1]^T$ and $\hat{s}_4 \neq s_4$, i.e., $\tilde{m}=M$. In this case, the path associated with $\hat{\vs}$ is in a totally different sub-tree from that associated with $\vs$. As a result, a large number of nodes are explored to correct $\hat{s}_4$ and find the optimal solution. The example in Fig. \ref{fig_tree_SESD} clearly shows that the search efficiency in SD significantly depends on the erroneous layer. Motivated by this, we propose a layer-ordering scheme such that errors are more likely to occur at low layers.
	
	In this scheme, the accuracy of the symbols in $\hat{\vs}$ are evaluated. For this purpose, we propose exploiting the difference between $\hat{\vs}^{[L]}$ and $\hat{\vs}$, which are the output of the last layer and the final solution of the FS-Net, respectively. We recall that $\hat{\vs}^{[L]} \in \setR^{M \times 1}$ can contain elements both inside and outside alphabet $\setA$, as observed from step 5 in Algorithm \ref{al_fsnet} and Fig. \ref{fig_phi}. In contrast, $\hat{\vs} = \mathcal{Q} \left( \hat{\vs}^{[L]} \right) \in \setA^M$. Let $e_m$ denote the distance between $\hat{s}_m$ and $\hat{s}_m^{[L]}$, i.e., $e_m = \abs{ \hat{s}_m^{[L]} - \hat{s}_m }$.
	For QAM signals, the distance between two neighboring real symbols is two. Furthermore, from Fig. \ref{fig_phi}, $\hat{s}_m^{[L]} \in [-1, 1]$ for QPSK, $\hat{s}_m^{[L]} \in [-3, 3]$ for 16-QAM, and $\hat{s}_m^{[L]} \in [-7, 7]$ for 64-QAM. Therefore, $0 \leq e_m \leq 1, m=1,2,\ldots,M$. It is observed that if $e_m \approx 0$, there is a high probability that the $m$th symbol in $\hat{\vs}$ is correctly approximated by the FS-Net, i.e., $s_m = \hat{s}_m$. In contrast, if $e_m \approx 1$, there is a high probability that $\hat{s}_m$ is an erroneous estimate, i.e., $s_m \neq \hat{s}_m$. Therefore, by examining the elements of $\ve = [ e_1, e_2, \ldots, e_M ]$, we can determine the layers with high probabilities of errors.
	
	Based on $\hat{\vs}^{[L]}$ and $\hat{\vs}$, $\ve$ is computed, and the layers are ordered in decreasing order of the elements of $\ve$ to increase the likelihood that the errors occur at the low layers. In other words, we rearrange layers such that the $i$th lowest layer in the tree is associated with the $i$th largest element of $\ve$. We note that ordering the layers is equivalent to ordering the elements of $\vx$, which requires the corresponding column ordering of $\mH$. Therefore, in the proposed layer-ordering scheme, the channel columns are also ordered in the decreasing order by the magnitude of the elements of $\ve$. 
	
	\textit{Example 1:} Consider a MIMO system with $M=8$, QPSK, and $\hat{\vs}^{[L]} = [0.1, -0.65, -0.2, 0.85,$ $0.25, 0.9, -0.3, 1]^T$, $\hat{\vs} = [1,-1,-1,1,1,1,-1,1]^T$. Then, $\ve = \abs{ \hat{\vs}^{[L]} - \hat{\vs} } = [0.9, 0.35, 0.8, 0.15,$ $ 0.75, 0.1, 0.7, 0]^T$, which implies that the layer order should be $\{1,3,5,7,2,4,6,8\}$. Consequently, the channel columns should be ordered as $\underline{\mH} = \left[ \vh_1, \vh_3, \vh_5, \vh_7, \vh_2, \vh_4, \vh_6, \vh_8 \right]$, where $\vh_m$ is the $m$th column of $\mH$.
	
	The layer ordering allows the errors to be corrected earlier, which further accelerates the shrinking of the sphere in the SD scheme. As a result, the final solution can be found with reduced complexity compared to the case when layer ordering is not applied. Our simulation results show that a significant complexity reduction is attained owing to the layer ordering, especially at low SNRs. In particular, the complexity of the FDL-SD scheme with layer ordering is almost constant w.r.t. the SNR, unlike the conventional FP-SD and SE-SD schemes.
	
	\subsection{FDL-SD Algorithm}
	
	\begin{algorithm}[t]
		\caption{FDL-SD algorithm}
		\label{al_FDL_SD}
		\begin{algorithmic}[1]
			\REQUIRE $\mH, \vy$.
			\ENSURE $\hat{\vs}_{SD}$.
			\STATE {Find $\hat{\vs}$ and $\hat{\vs}^{[L]}$ based on Algorithm \ref{al_fsnet}.}
			\STATE {Obtain $\ve = \left[e_1, e_2, \ldots, e_M\right]^T$, where $e_m = \abs{ \hat{s}_m - \hat{s}_m^{[L]} }$.}
			\STATE Order the channel columns in decreasing order of the elements of $\ve$ to obtain $\underline{\mH}$.
			\STATE Perform QR decomposition of $\underline{\mH}$ to obtain $\mQ_1, \mQ_2, \mR$.
			\STATE Set $m=M$, $\vz = \mQ_1^T \vy$, $d^2 = \min \{\alpha N_r \sigma_n^2, \phi(\hat{\vs})\}$, $d_M^2 = d^2 - \norm{\mQ_2^T \vy}^2$.
			
			\STATE Compute $z_{m|m+1}$, $d_m^2$, $LB_m$, and $UB_m$ based on \eqref{bound_m}--\eqref{sigma_m}.
			
			\STATE Obtain $\mathcal{O}^{\text{FDL}}_m$ based on \eqref{order_FS}.
			\IF {$\mathcal{O}^{\text{FDL}}_m$ is empty}
			\STATE $m \leftarrow m+1$
			\ELSE
			\STATE Set $x_m$ to the first element in $\mathcal{O}^{\text{FDL}}_m$.
			\ENDIF 
			
			\IF {$m=1$}
			\STATE $\phi(\vx) = d_M^2 - d_1^2 + (z_1 - r_{1,1} x_1)^2$
			\IF {$\phi(\vx) \leq d_M^2$}
			\STATE Update $d_M^2 = \phi(\vx)$ and $\hat{\vs}_{SD} = \vx$.
			\STATE Remove the first element of $\mathcal{O}^{\text{FDL}}_m$ and go to step 8.
			\ENDIF 
			\ELSE
			\STATE Set $m \leftarrow m-1$ and go to step 6.
			\ENDIF 
		\end{algorithmic}
	\end{algorithm}
	
	The FDL-SD algorithm is summarized in Algorithm \ref{al_FDL_SD}. In step 1, the FS-Net is employed to obtain $\hat{\vs}$ and $\hat{\vs}^{[L]}$, which is then used in steps 2 and 3 for layer ordering {and in step 5 to predetermine the radius}. In the remaining steps, the common search process of SD is conducted to obtain $\hat{\vs}_{SD}$. Note that in step 7, all the symbols belonging to the interval $[LB_m, UB_m]$ are ordered by increasing distance from $\hat{s}_m$, as given in \eqref{order_FS}. Performing this operation for every layer allows the candidates to be examined by their increasing distance to the FS-Net's solution $\hat{\vs}$. The remaining steps follow the well-known search procedure of SD \cite{hassibi2005sphere}.
	
	Compared to the existing DL-aided SD schemes in \cite{askri2019dnn, mohammadkarimi2018deep,weon2020learning}, the proposed FDL-SD algorithm is advantageous in the following aspects:
	\begin{itemize}
		\item The application of DL in this scheme is to generate a highly reliable candidate $\hat{\vs}$. We note that in this employment, the DNN, i.e., the FS-Net, can be trained without performing the conventional SD scheme, as will be further discussed in Section VI. In contrast, in \cite{askri2019dnn, mohammadkarimi2018deep,weon2020learning}, DL is applied to predict the radius, and its training labels are obtained by performing the conventional SD scheme. This requires considerable time and computational resources. For example, to train the DNN in \cite{weon2020learning} for a $16 \times 16$ MIMO system with QPSK, 100,000 samples are used, requiring performing the conventional SD 100,000 times to collect the same number of desired radii for training, whereas that number required in \cite{mohammadkarimi2018deep} for a $10 \times 10$ MIMO system with 16-QAM is 360,000. This computational burden in the training phase of the existing DL-aided SD schemes is non-negligible, even for offline processing.
		
		\item The proposed scheme does not require optimizing the initial radius, in contrast to \cite{askri2019dnn, mohammadkarimi2018deep,weon2020learning}, because the initial sphere is predetermined based on $\hat{\vs}$, {as shown in step 5 of Algorithm \ref{al_FDL_SD}}. Note that in the conventional SD, if the radius is initialized to a small value, it is possible that there will be no point inside the sphere. In this case, the search needs to restart with a larger radius, resulting in redundant complexity. In contrast, in the FDL-SD scheme, starting with $\hat{\vs}$ guarantees that there is always at least one point inside the sphere, which is nothing but $\hat{\vs}$. Furthermore, because $\hat{\vs}$ has high accuracy, the number of points inside the sphere is typically small.
		
		\item In the proposed FDL-SD, the search efficiency is improved, thus providing significant complexity reduction. Despite that, the ordering schemes in the FDL-SD do not affect the radius or terminate the search early, in contrast to \cite{askri2019dnn, mohammadkarimi2018deep,weon2020learning}. Therefore, the BER performance of the conventional SD is totally preserved in the proposed FDL-SD scheme. We will further justify this with the simulation results in Section VI.
	\end{itemize}

	\section{Proposed FDL-KSD Scheme}
	
	It is intuitive from the tree-like model shown in Fig. \ref{fig_tree_model}(b) that there are $\abs{\mathcal{A}}^M$ complete paths representing all the possible candidates for the optimal solution, where $\abs{\mathcal{A}}=2$ and $M=4$ for the example in Fig. \ref{fig_tree_model}(b). In a large MIMO system with a high-order modulation scheme, i.e., when $M$ and $\abs{\mathcal{A}}$ are large, the number of paths becomes very large. Therefore, in the KSD, instead of examining all the available paths, only the $K$ best paths are selected in each layer for further extension to the lower layer, while the others are pruned early to reduce complexity. For the selection of the $K$ best paths, each path is evaluated based on its metric. Specifically, in layer $m$, if the $k$th path extends to a node $x_i$, its metric is given by
	\begin{align*}
	\phi_m^{(k,i)} = \phi_{m+1}^{(k)} + \left(z_m - \sum_{i=m}^{M} r_{m,i} x_i \right) ^2, \nbthis \label{path_metric}
	\end{align*}
	with $\phi_{M+1}^{(k)} = 0, \forall k$. Then, only a subset of $K$ paths with the smallest metrics $\{\phi_{m}^{(1)}, \ldots, \phi_{m}^{(K)}\}$ are selected for further extension. In the lowest layer, the best path with the smallest metric is concluded to be the final solution. In this study, to further optimize the KSD scheme in terms of both complexity and performance, we propose the FDL-KSD scheme with early rejection and layer ordering, which is presented in the following subsection.

	\subsection{Basic ideas: Early Rejection and Layer Ordering}

	\textit{Early rejection:} The idea of early rejection is that, given the output  $\hat{\vs}$ of the FS-Net, a candidate that is worse than $\hat{\vs}$ cannot be the optimal solution, and it can be rejected early from the examination process. This definitely results in complexity reduction without any performance loss. To apply this idea to the KSD, among the $K$ chosen paths in each layer of the KSD scheme, the paths with metrics larger than $d^2 = \min \{\alpha N_r \sigma_n^2, \phi(\hat{\vs})\}$ are pruned early because their corresponding candidates are worse than $\hat{\vs}$ or outside the sphere. It is possible that all the $K$ paths are pruned in a layer if all of them are worse than $\hat{\vs}$. In this case, there is no path for further extension. Hence, the examination process is terminated early, and $\hat{\vs}$ is concluded to be the final solution. It is observed that in this early rejection approach, the paths with the metrics larger than $\phi(\hat{\vs})$ are pruned, thus the final solution is the best one between that attained by the conventional KSD and the FS-Net-based solution. Therefore, besides providing complexity reduction, this scheme also attains performance improvement w.r.t. the conventional KSD.

	\textit{Layer ordering:} One potential problem of the KSD is that the optimal solution can be rejected before the lowest layer is reached, causing its performance loss w.r.t. the sequential SD. An approach to mitigate the unexpected early rejection of the optimal solution is to apply the layer-ordering scheme proposed in Section IV-B. Specifically, it is observed from \eqref{path_metric} that if the elements of a candidate $\vx$ are ordered such that the ones in higher layers are more reliable than those in lower layers, then the best path is more likely to have small metrics at high layers. As a result, the chance that it is early pruned is reduced. Therefore, we propose applying the layer-ordering scheme proposed in Section IV-B to the FDL-KSD scheme for performance improvement.
	
	\subsection{FDL-KSD Algorithm}
	
	\begin{algorithm}[t]
		\caption{FDL-KSD algorithm}
		\label{al_FDL_KSD}
		\begin{algorithmic}[1]
			\REQUIRE $\mH, \vy, K$.
			\ENSURE $\hat{\vs}_{KSD}$.
			\STATE {Find $\hat{\vs}$ and $\hat{\vs}^{[L]}$ based on Algorithm \ref{al_fsnet}.}
			\STATE {Obtain $\ve = \left[e_1, e_2, \ldots, e_M\right]^T$, where $e_m = \abs{ \hat{s}_m - \hat{s}_m^{[L]} }$.}
			\STATE Order the channel columns in decreasing order of the elements of $\ve$ to obtain $\underline{\mH}$.
			\STATE Perform QR decomposition of $\underline{\mH}$ to obtain $\mQ_1, \mQ_2, \mR$.
			\STATE {$\vz = \mQ_1^T \vy$, $d^2 = \min \{\alpha N_r \sigma_n^2, \phi(\hat{\vs})\}$}
			\FOR {$m = M \rightarrow 1$}
			\STATE Determine the $K$ best paths associated with the $K$ smallest metrics $\{\phi_{m}^{(1)}, \ldots, \phi_{m}^{(K)}\}$.
			\STATE Prune the paths that have metrics larger than $d^2$ early.
			\IF {all paths have been pruned}
			\STATE Terminate the search early.
			\ENDIF
			
			\STATE Save the survival paths for further extension.
			\ENDFOR
			
			\IF {there is no survival path}
			\STATE Set $\hat{\vs}_{KSD} = \hat{\vs}$.
			\ELSE
			\STATE Set $\hat{\vs}_{KSD}$ to the survival candidate corresponding to the path with the smallest metric.
			\ENDIF
		\end{algorithmic}
	\end{algorithm}
	
	The proposed FDL-KSD scheme is summarized in Algorithm \ref{al_FDL_KSD}. In steps 1--3, layer ordering is performed. The $K$ best paths are selected in step 7, and a subset of them with metrics larger than $d^2 = \min \{\alpha N_r \sigma_n^2, \phi(\hat{\vs})\}$ are pruned early in step 8. In the case where all the paths are pruned, the path examination and extension process is terminated early in step 10, and the FS-Net-based solution $\hat{\vs}$ is concluded to be the final solution, as shown in step 15. In contrast, if early termination does not occur, the search continues until the lowest layer is reached, at which the final solution $\hat{\vs}_{KSD}$ is set to be the best candidate among the surviving ones, as in step 17.
	
	We further discuss the properties of $K$, i.e., the number of survival paths in the proposed FDL-KSD scheme. It can be seen that in the proposed scheme, the number of actual survival paths is dynamic, whereas it is fixed to $K$ in the conventional KSD scheme. Letting $K_m$ be the number of survival paths in the $m$th layer, we have $K_m \leq K, \forall m$, which is clear from step 8 of Algorithm \ref{al_FDL_KSD}. Furthermore, it is observed in \eqref{path_metric} that the paths' metrics increase with $m$. As a result, more paths have metrics exceeding $d^2$ as $m$ increases. Consequently, the number of survival paths becomes smaller as the search goes downward to lower layers, i.e., $K_1 \geq K_2 \geq \ldots \geq K_M$.
	
	These properties make the design of the FDL-KSD scheme much easier than that of the conventional KSD scheme. We first note one challenge in the conventional KSD, which is to choose the optimal value for $K$. Specifically, if a large $K$ is set, many candidates are examined, resulting in high complexity. In this case, the complexity reduction of KSD w.r.t. the conventional SD is not guaranteed. In contrast, a small $K$ leads to significant performance loss because there is a high probability that the optimal path is pruned before the lowest layer is reached. It is possible to use dynamic $K$, i.e., to set different values of $K$ for different layers. However, optimizing multiple values of $\{K_1, K_2, \ldots, K_M\}$ becomes problematic, as $M$ is large in large MIMO systems. In the proposed FDL-KSD scheme, $K_m$ is already dynamic. Furthermore, because $K_m$ is adjusted in step 8 of Algorithm \ref{al_FDL_KSD}, we only need to set $K$ to a sufficiently large value to guarantee near-optimal performance, and unpromising paths are automatically rejected by the FDL-KSD scheme.

	\section{Simulation Results}
	\label{sec_sim_result}
	
	In this section, we numerically evaluate the BER performance and computational complexities of the proposed FDL-SD and FDL-KSD schemes. The computational complexity of an algorithm is calculated as the total number of additions and multiplications required for online signal detection. Because the training phase of a DL model can be performed offline, the computational complexity in this phase is ignored. In our simulations, each channel coefficient is assumed to be an i.i.d. zero-mean complex Gaussian random variable with a variance of $1/2$ per dimension. SNR is defined as the ratio of the average transmit power to the noise power, i.e., SNR $=N_t\smt / \smn$. 
	
	We consider the following schemes for comparison:
		\begin{itemize}
			\item Conventional SD schemes: FP-SD, SE-SD, and KSD.
			\item Existing DL-aided detection schemes: MR-DL-SD \cite{mohammadkarimi2018deep}, DPP-SD \cite{weon2020learning}, and FS-Net \cite{nguyen2019deep}.
			\item SD with the ordered SIC (OSIC)-based initial solution (OSIC-SD).
			\item TS algorithm aided by DL (DL-TS) \cite{nguyen2019deep}.
			\item OAMP detection method \cite{he2018model}.
		\end{itemize}
		More specifically, we present the BER performance and complexity reduction attained by the proposed FDL-SD w.r.t. the conventional FP-SD and SE-SD, which is shown to be much more significant than that achieved by the existing MR-DL-SD and DPP-SD schemes. Furthermore, we also show the BER performance of the FS-Net to demonstrate the gains of incorporating FS-Net with SD in the proposed FDL-SD and FDL-KSD schemes. Similar observations are noted from the comparison between FDL-KSD and conventional KSD and FS-Net. To justify the efficiency of using the FS-Net-based initial solution, we demonstrate the performance and complexity of the OSIC-SD. Furthermore, we compare the FDL-SD and FDL-KSD to the OAMP \cite{he2018model} scheme. For the OAMP scheme, we have performed simulations to select the number of iterations, denoted by $T$, to ensure convergence. Based on this, we set $T = 4$ for systems with QPSK and $T=8$ for systems with 16-QAM and 64-QAM. For the proposed FDL-SD scheme, we also show the BER performance and computational complexity when only candidate ordering is applied, which allows us to compare the efficiency of order $\mathcal{O}^{\text{FDL}}_m$ in \eqref{order_FS} proposed for the FDL-SD scheme and $\mathcal{O}^{\text{SE}}_m$ in \eqref{order_SE} employed in the conventional SE-SD scheme. Finally, we compare the proposed FDL-SD to the DL-TS \cite{nguyen2019deep} in terms of both the BER performance and complexity to show that, although these two schemes both leverage the FS-Net, the former is more efficient than the latter in both aspects.

	\begin{table*}[t]
		\renewcommand{\arraystretch}{1.4}
		\caption{Architectures and complexities of the DNNs used in the MR-DL-SD, DDP-SD, and the proposed FDL-SD schemes for a $16 \times 16$ MIMO system with QPSK.}
		\label{tab_FCDNN}
		\centering
		\begin{tabular}{|c|c|c|c|c|}
			\hline
			DNNs   & \makecell{No. of \\input nodes} & \makecell{No. of hidden nodes \\$\times$ No. of hidden layers} & \makecell{No. of \\output nodes} & \makecell{Complexity \\ (operations)} \\
			\hline
			\hline
			
			FC-DNN in the MR-DL-SD  & $544$  & $128 \times 1$ & $4$ &  70276  \\
			\hline
			
			FC-DNN in the DPP-SD  & $34$  & $40 \times 1$ & $4$ & 1564  \\
			\hline
			
			\makecell{FS-Net in the FDL-SD, FDL-KSD, and DL-TS}  & $64$  & $64 \times 10$ & $32$ & $88608$  \\
			\hline
		\end{tabular}
	\end{table*}
	
	\begin{figure*}[t]
		\centering
		\includegraphics[scale=0.55]{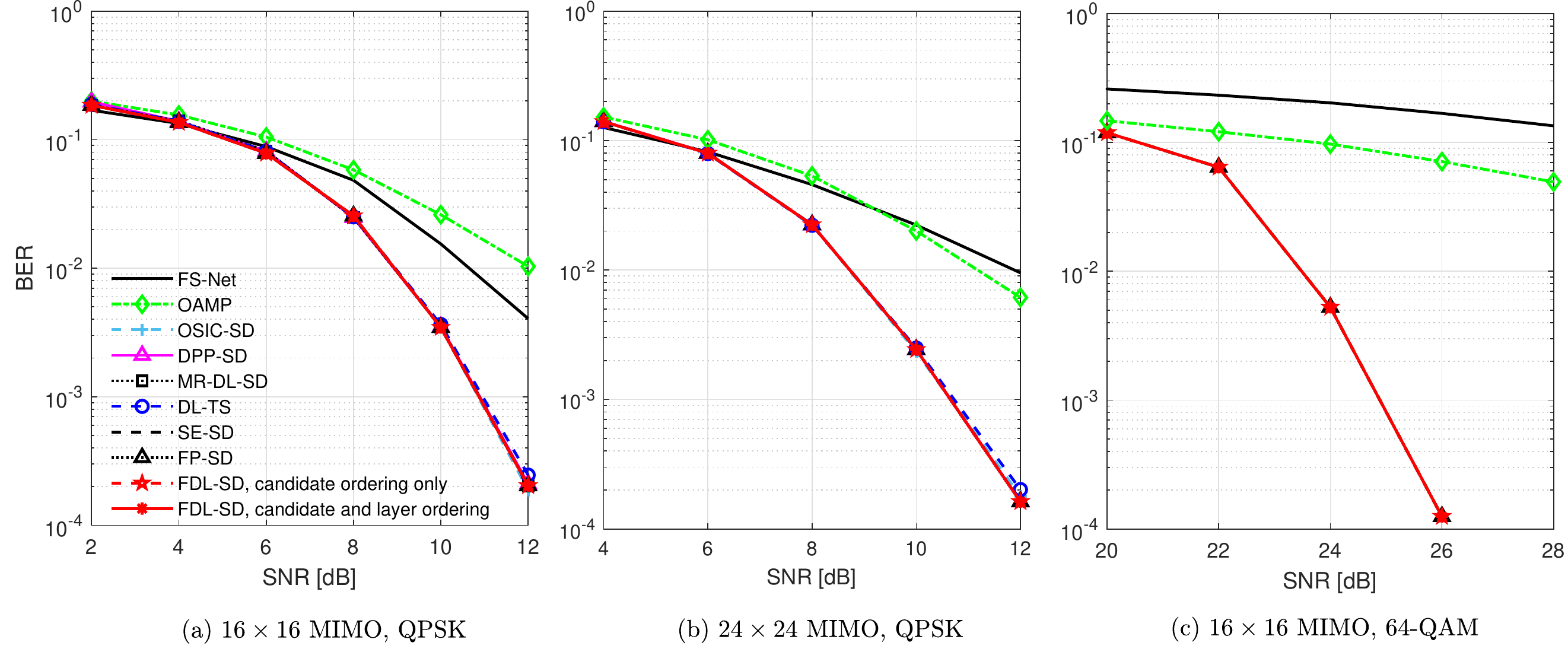}
		\caption{BER performance of the proposed FDL-SD scheme compared to those of the conventional FP-SD, SE-SD, MR-DL-SD, DPP-SD, DL-TS, OSIC-SD, {and OAMP} for $16 \times 16$ MIMO with QPSK and $L=10$, $24 \times 24$ MIMO with QPSK and  $L = 12$, {and $16 \times 16$ MIMO with 64-QAM and $L = 15$}.}
		\label{fig_ber_SD}
	\end{figure*}
	
	\begin{figure*}[t]
		\centering
		\includegraphics[scale=0.52]{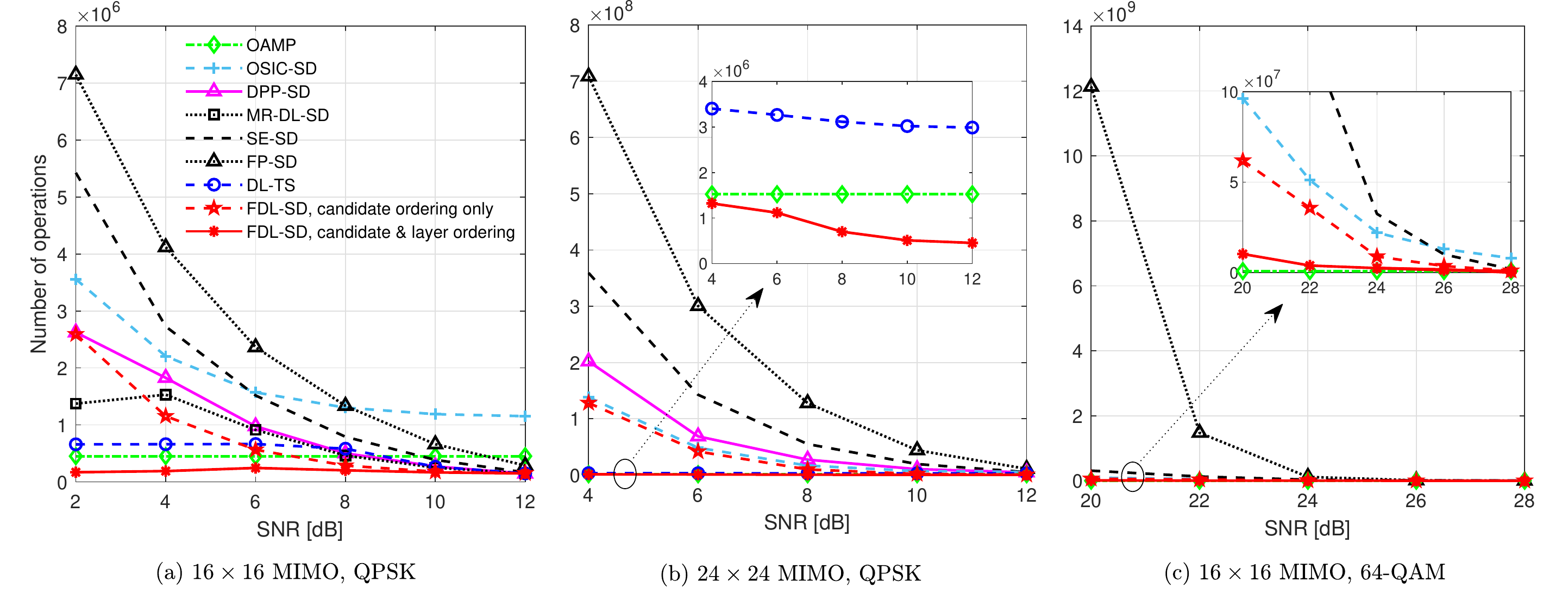}
		\caption{Complexity of of the proposed FDL-SD scheme compared to those of the conventional FP-SD, SE-SD, MR-DL-SD, DPP-SD, DL-TS, OSIC-SD, {and OAMP} for $16 \times 16$ MIMO with QPSK and $L=10$, $24 \times 24$ MIMO with QPSK and  $L = 12$, {and $16 \times 16$ MIMO with 64-QAM and $L = 15$}.}
		\label{fig_comp_SD}
	\end{figure*}

	\subsection{Training DNNs}
	
	The hardware and software used for implementing and training the DNNs are as follows. The FS-Net is implemented by using Python with the TensorFlow library \cite{abadi2016tensorflow}. In contrast, the FC-DNNs in the MR-DL-SD and DPP-SD are implemented using the DL Toolbox of MATLAB 2019a, as done in \cite{mohammadkarimi2018deep} and \cite{weon2020learning}. All the considered DNNs, i.e., the FS-Net and FC-DNNs, are trained by the Adam optimizer \cite{kingma2014adam,rumelhart1988learning, bottou2010large} with decaying and starting learning rates of $0.97$ and $0.001$, respectively. The FC-DNNs used for the MR-DL-SD and DPP-SD are trained for 100,000 samples, as in \cite{weon2020learning}. In contrast, we train the FS-Net for 10,000 epochs with batch sizes of 2,000 samples, as in \cite{nguyen2019deep}. For each sample, $\vs, \mH$, and $\vy$ are independently generated from \eqref{real SM}.
	
	As discussed in Section I, one of the significant differences between the proposed and existing DL-aided SD schemes lies in the training phase. In the existing DL-aided SD schemes, including the SR-DL-SD, MR-DL-SD, and DPP-SD, the FC-DNNs are trained with the following loss function
	\begin{align*}
		\mathcal{L} \left(\vd^{(i)}, \hat{\vd}^{(i)} \right) = \frac{1}{D} \sum_{i=1}^{D} \norm{\vd^{(i)} - \hat{\vd}^{(i)}}^2,
	\end{align*}
	where $D$ is the number of training data samples, and $\vd^{(i)}$ and $\hat{\vd}^{(i)}$ are the label and output vectors of the DNNs, which represent the radii associated with the $i$th data sample \cite{mohammadkarimi2018deep,weon2020learning}. The training labels, i.e., $\{\vd^{(1)}, \ldots, \vd^{(D)}\}$, are obtained by performing the conventional SD scheme for $D$ times, where $D = \{\text{100,000}, \hspace{0.05cm} \text{360,000}\}$ for the DPP-SD and MR-DL-SD schemes, respectively \cite{mohammadkarimi2018deep, weon2020learning}. It is well known that the conventional SD is computationally prohibitive for large MIMO systems. Therefore, huge amounts of computational resources and time are required to collect a huge training data set in the existing DL-aided SD schemes.
	
	In contrast, in the proposed application of DL to SD, the FS-Net is employed to generate $\hat{\vs}$. The FS-Net is trained with the loss function \eqref{loss_dscnet} \cite{nguyen2019deep}. As the training labels $\vs$ of the FS-Net are generated randomly, the conventional SD does not need to be performed to generate the training labels as done in the existing DL-aided SD schemes.

	\subsection{BER Performance and Computational Complexity of the Proposed FDL-SD Algorithm}

	We first note that the structures and complexities of the DNNs employed in the compared schemes are different, as illustrated in Table \ref{tab_FCDNN} for a $16 \times 16$ MIMO system with QPSK. The DNNs used in the MR-DL-SD and DPP-SD schemes have well-known fully-connected architectures, whose complexity can be calculated based on the network connections. In contrast, the complexity of the FS-Net is computed based on \eqref{comp_fsnet}. In our simulation results, the overall complexity of each considered scheme is computed as the sum of the complexity required in the DNNs, presented in Table \ref{tab_FCDNN}, and that required to perform {QR decomposition and the search process in the} algorithms themselves. The simulation parameters for the MR-DL-SD, DPP-SD, and DL-TS schemes are listed in Table \ref{tabl_sim_params}, which are set based on the corresponding prior works. It is observed that an advantage of the proposed FDL-SD scheme is that it does not require optimizing any design parameters, as done in the DL-aided detection algorithms in Table \ref{tabl_sim_params}.
	
	\begin{table*}[t]
		\renewcommand{\arraystretch}{1.4}
		\caption{Simulation parameters for the MR-DL-SD, DPP-SD, and DL-TS schemes, where $\lambda_1, \lambda_2$ are the optimized design parameters of the DPP-SD scheme, $\mathcal{I}$ is the maximum number of search iterations, $\mu$ is an optimized design parameter, and $\epsilon$ is the cutoff factor for early termination in the DL-TS scheme.}
		\label{tabl_sim_params}
		\centering
		\begin{tabular}{|c|c|}
			\hline
			Schemes  & Parameters \\
			\hline
			\hline
			
			MR-DL-SD  & Number of predicted radii: 4 \\
			\hline
			
			DPP-SD  & $\lambda_1 = \{ 1.1, 1.2 \ldots, 1.6\}$ for SNR $= \{2,4,\ldots, 12\}$, respectively, $\lambda_2 = \lambda_1 + 0.1$ \\
			\hline
			
			DL-TS & $\mathcal{I} = \{ 400, 700 \}, \mu = \{7, 8\}$ for $16 \times 16$ MIMO and $24 \times 24$ MIMO systems, respectively, and $\epsilon = 0.4$ \\
			\hline
		\end{tabular}
	\end{table*}
	
	\begin{figure*}[t]
		\centering
		\includegraphics[scale=0.55]{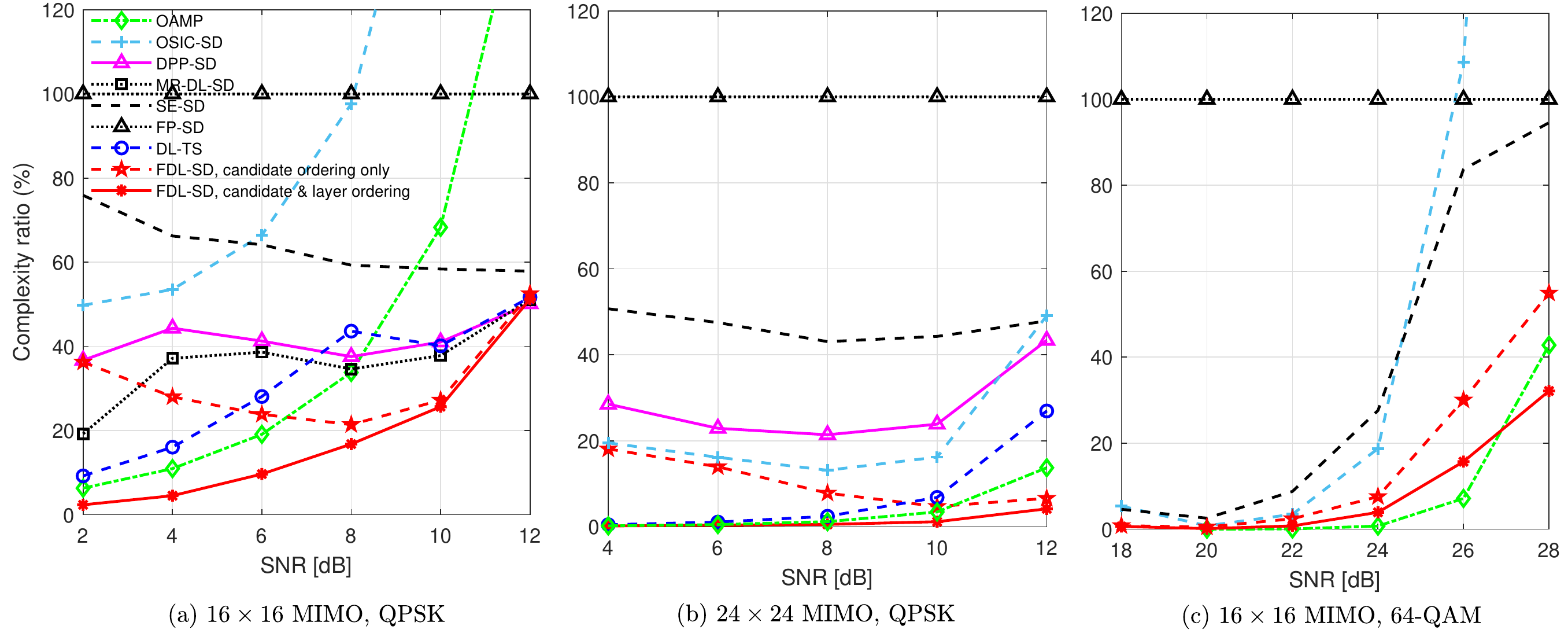}
		\caption{Complexity ratio of the proposed FDL-SD scheme compared to those of the conventional FP-SD, SE-SD, MR-DL-SD, DPP-SD, DL-TS, OSIC-SD, {and OAMP} for $16 \times 16$ MIMO with QPSK and $L=10$, $24 \times 24$ MIMO with QPSK and  $L = 12$, {and $16 \times 16$ MIMO with 64-QAM and $L = 15$}.}
		\label{fig_comp_ratio_SD}
	\end{figure*}

	\begin{figure}[t]
		\centering
		\subfigure[$16 \times 16$ MIMO with QPSK]
		{
			\includegraphics[scale=0.55]{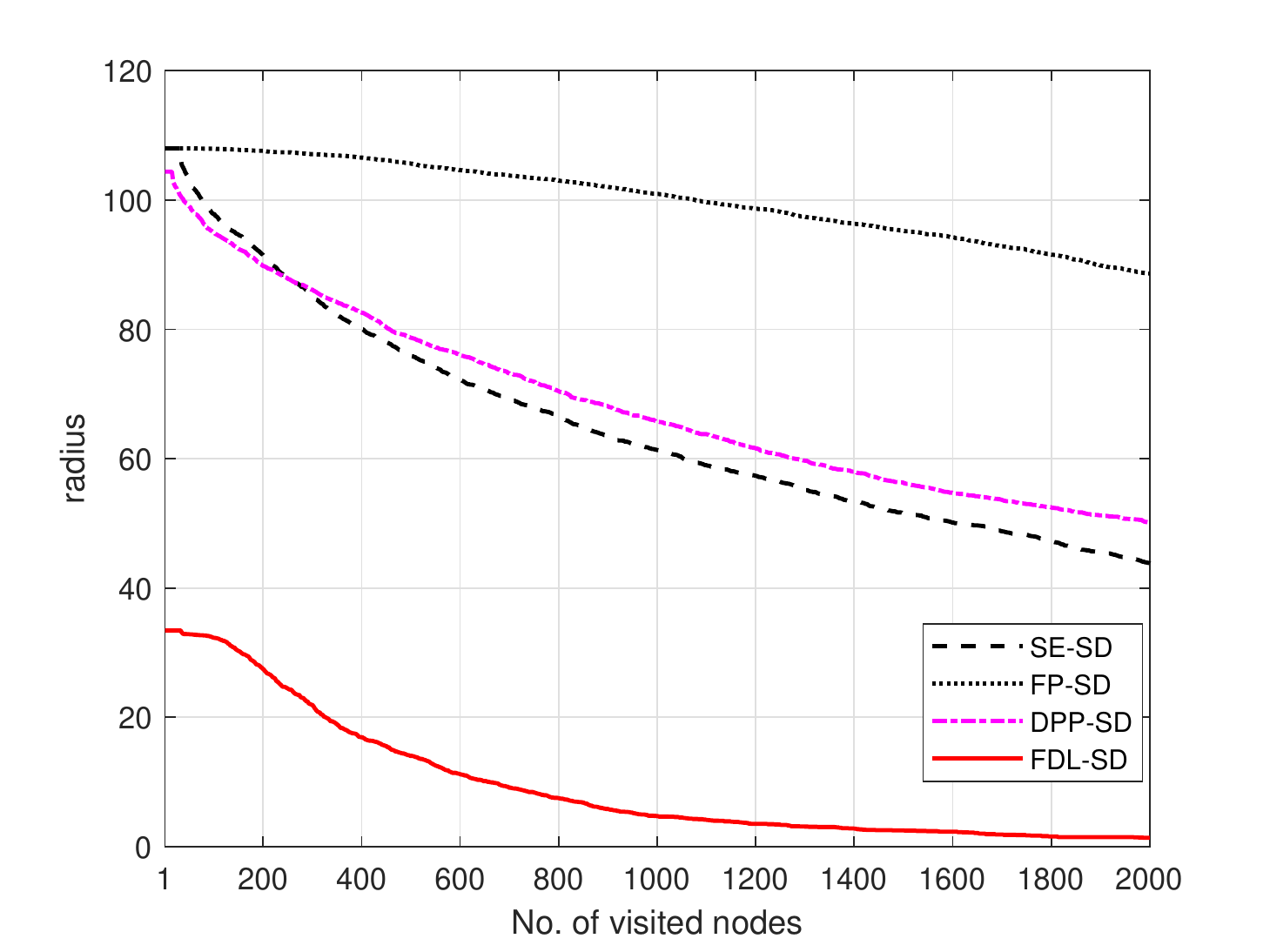}
			\label{conv_16QPSK}		}
		\subfigure[$24 \times 24$ MIMO with QPSK]
		{
			\includegraphics[scale=0.55]{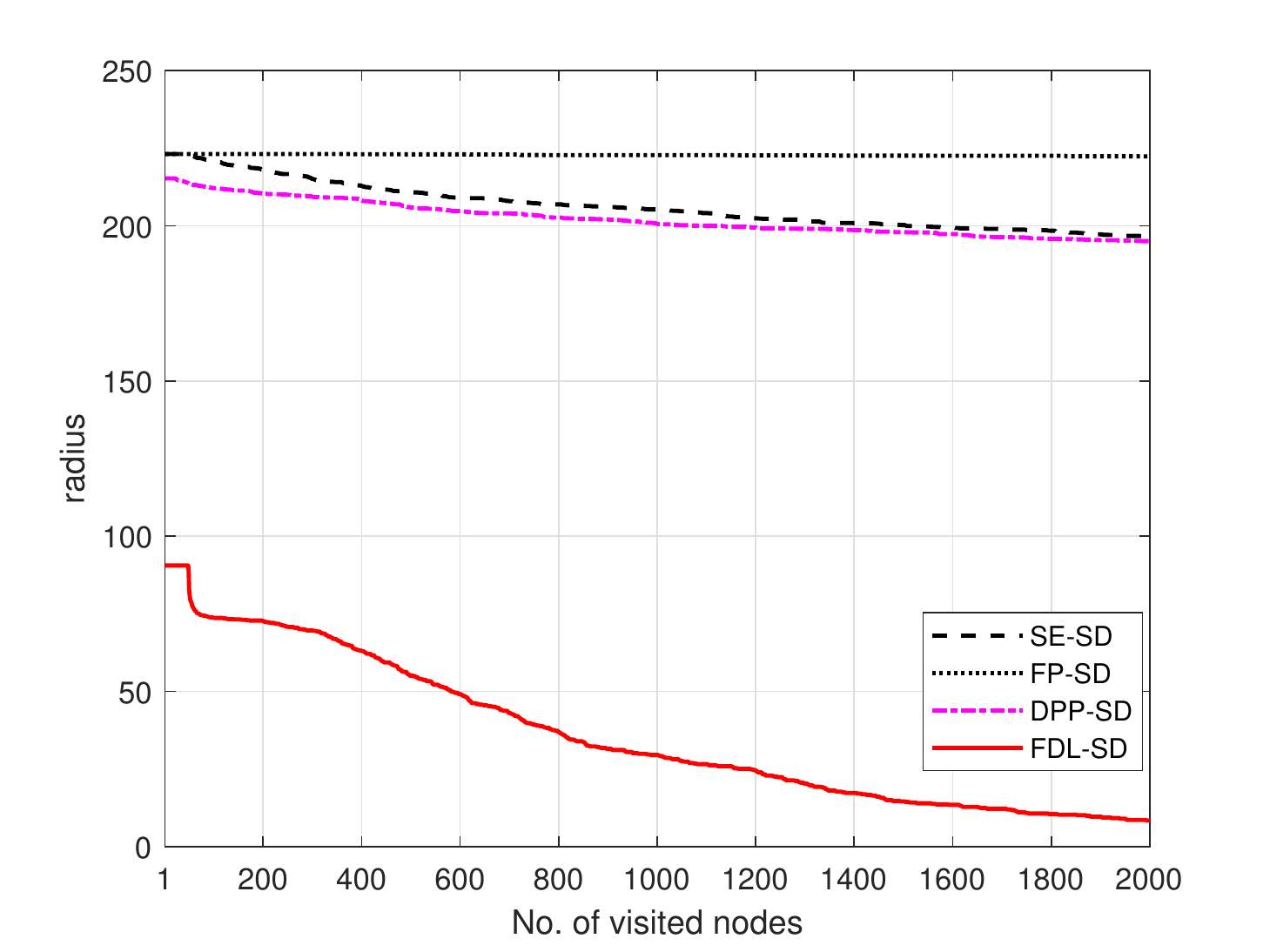}
			\label{conv_24QPSK}
		}
		\caption{{Convergence of the proposed FDL-SD scheme compared to those of the conventional FP-SD, SE-SD, and DPP-SD.}}
		\label{fig_conv}
	\end{figure}
	
	{In Fig. \ref{fig_ber_SD}, we show the BER performance of the schemes listed earlier for $16 \times 16$ and $24 \times 24$ MIMO systems, both with QPSK, and $L=\{10,12\}$, respectively, and a $16 \times 16$ MIMO system with 64-QAM and $L=15$. We note that the results for the MR-DL-SD and DPP-SD schemes are not presented in Figs.\ \ref{fig_ber_SD}(b) and (c) because it takes an extremely long time to collect the desired radii to train the corresponding DNNs. It is seen from Fig. \ref{fig_ber_SD} that, except for the FS-Net and OAMP schemes, the compared schemes, including the FDL-SD, MR-DL-SD, DPP-SD, DL-TS, FP-SD, and SE-SD, have approximately the same BER performance, which is near-optimal. In particular, the proposed FDL-SD schemes completely preserve the performance of the conventional FP-SD and SE-SD because the incorporation with the FS-Net solution does not affect the final solution. The OAMP scheme performs far worse than the considered SD schemes, which agrees with the observations in \cite{he2018model}. However, in a larger system with a higher modulation order, it achieves better performance than FS-Net, as observed in Figs.\ \ref{fig_ber_SD}(b) and (c).}
	
	In Fig. \ref{fig_comp_SD}, we compare the proposed FDL-SD scheme to the conventional FP-SD, SE-SD, MR-DL-SD, DPP-SD, {OSIC-SD}, DL-TS, {and OAMP} schemes in terms of computational complexity. To ensure that the compared schemes have approximately the same BER performance, the simulation parameters in Fig. \ref{fig_comp_SD} are assumed to be the same as those in Fig. \ref{fig_ber_SD}. In Fig. \ref{fig_comp_SD}, the complexity reduction gains of the considered schemes are difficult to compare at high SNRs. Therefore, we show their complexity ratios w.r.t. the complexity of the conventional FP-SD in Fig. \ref{fig_comp_ratio_SD}. In other words, the complexity of all schemes {is} normalized by that of the FP-SD. From Figs. \ref{fig_comp_SD} and \ref{fig_comp_ratio_SD}, the following observations are noted:
	\begin{itemize}
		\item It is clear from Fig. \ref{fig_comp_SD} that the complexities of the conventional FP-SD, SE-SD, {OSIC-SD}, and the existing DL-aided SD schemes, including the MR-DL-SD and DPP-SD, significantly depend on SNRs. In contrast, that of the proposed FDL-SD scheme is relatively stable with SNRs. 
		
		\item In Fig. \ref{fig_comp_ratio_SD}, among the improved SD schemes, the proposed FDL-SD achieves the most significant complexity reduction w.r.t. the conventional FP-SD scheme. Specifically, in the $16 \times 16$ MIMO system, for SNR $\leq 6$ dB, the complexity reduction ratios of the proposed FDL-SD are higher than $90\%$, while those of the MR-DL-SD and DPP-SD are only around $60\%$. At SNR $=12$ dB, the DL-aided SD schemes, including MR-DL-SD, DPP-SD, and FDL-SD, have approximately the same complexity. In the $24 \times 24$ MIMO system, the complexity reduction ratio of the FDL-SD with both candidate and layer ordering is $95\% - 98\%$, which is much higher than $55\% - 80\%$ for the DPP-SD scheme.
		
		\item Furthermore, by comparing the complexity ratios of the FDL-SD scheme in Fig. \ref{fig_comp_ratio_SD}, it can be observed that this scheme achieves more significant complexity reduction in a larger MIMO system. Specifically, in the $16 \times 16$ MIMO system, its complexity reduction ratio w.r.t. the conventional FP-SD is only around $50\% - 97\%$. In contrast, that in the $24 \times 24$ MIMO system with QPSK is $95\% - 98\%$. The reason for this improvement is that, in large MIMO systems, the complexity of the SD algorithm significantly dominates that of the FS-Net and becomes almost the same as the overall complexity. Therefore, the complexity required in the FS-Net has almost no effect on the complexity of the FDL-SD scheme. This observation demonstrates that the proposed FDL-SD scheme is suitable for large MIMO systems.
		
		\item Notably, the proposed FDL-SD has considerably lower complexity than the DL-TS scheme although the conventional SD requires higher complexity than the TS detector \cite{nguyen2019qr, nguyen2019groupwise}. {This confirms that in the considered scenarios, the application of DL makes SD a more computationally efficient detection scheme than the TS.} Furthermore, although the OSIC-SD scheme attains complexity reduction at low and moderate SNRs, this is not guaranteed at high SNRs. This is because high complexity is required to obtain the OSIC solution, whereas that to perform SD at high SNRs is relatively low; thus, the complexity reduction achieved by using the OSIC solution cannot compensate for the complexity increase of the OSIC-SD scheme at high SNRs.
		\item Compared to the conventional FP-SD, the SE-SD and the proposed FDL-SD scheme with candidate ordering only are similar in the sense that symbols are ordered in each layer based on $\mathcal{O}^{\text{FDL}}_m$ and $\mathcal{O}^{\text{SE}}_m$, respectively. However, it can be clearly seen in Fig. \ref{fig_comp_ratio_SD} that the order $\mathcal{O}^{\text{FDL}}_m$ obtained based on the FS-Net's output is considerably better than the $\mathcal{O}^{\text{SE}}_m$ used in the conventional SE-SD. Specifically, in the $24 \times 24$ MIMO system, the proposed FDL-SD with candidate ordering based on $\mathcal{O}^{\text{FDL}}_m$ achieves $80\% - 95\%$ complexity reduction w.r.t. the conventional FP-SD, while that achieved by the SE-SD with $\mathcal{O}^{\text{SE}}_m$ is only around $50\%$, as seen in Fig. \ref{fig_ber_SD}.
		
		\item {The FS-Net performs worse for higher-order modulations, such as 64-QAM. However, the complexity reduction ratio of the FDL-SD compared to that of the FP-SD is still significant, $70\%$--$99\%$ in Fig. \ref{fig_comp_ratio_SD}(c). It can be observed from Figs.\ \ref{fig_comp_SD} and \ref{fig_comp_ratio_SD} that the OAMP scheme has relatively low complexity. In particular, at low SNRs, its complexity is much lower than that of most of the compared schemes, except for the proposed FDL-SD. However, it is noted that this does not guarantee near-optimal performance, as shown in Fig.\ \ref{fig_ber_SD}.}
	\end{itemize}

	{To explain the complexity reduction of the proposed FDL-SD scheme, we further investigate its convergence compared to those of the FP-SD, SE-SD, and DPP-SD schemes in Fig. \ref{fig_conv} for $16 \times 16$ and $24 \times 24$ MIMO systems with QPSK. Based on the description of the SD scheme in Section IV, the convergence of the SD schemes can be evaluated by the number of visited nodes, each associated with an examined candidate symbol. Specifically, the number of visited nodes is equal to the number of iterations that the SD schemes perform until convergence. For example, in the proposed FDL-SD algorithm, this iterative process is conducted during steps 6--21 of Algorithm \ref{al_FDL_SD}, which is similar to the conventional SD schemes \cite{hassibi2005sphere}. This iterative search process terminates when the sphere stops shrinking, i.e., when the radius stops decreasing and reaches convergence. In Fig. \ref{fig_conv}, we show the convergences of the considered schemes for $2000$ iterations/visited nodes. It is observed that the FDL-SD scheme reaches convergence much earlier than the other compared schemes.} In summary, it is clear from Figs. \ref{fig_ber_SD}--\ref{fig_conv} that the proposed FDL-SD scheme has no performance loss w.r.t. the conventional SD, whereas it attains the most significant complexity reduction among the compared schemes.
	
	\subsection{BER Performance and Computational Complexity of the Proposed FDL-KSD Scheme}
	
	\begin{figure}[t]
		%\vspace{-1cm}
		\centering
		\includegraphics[scale=0.55]{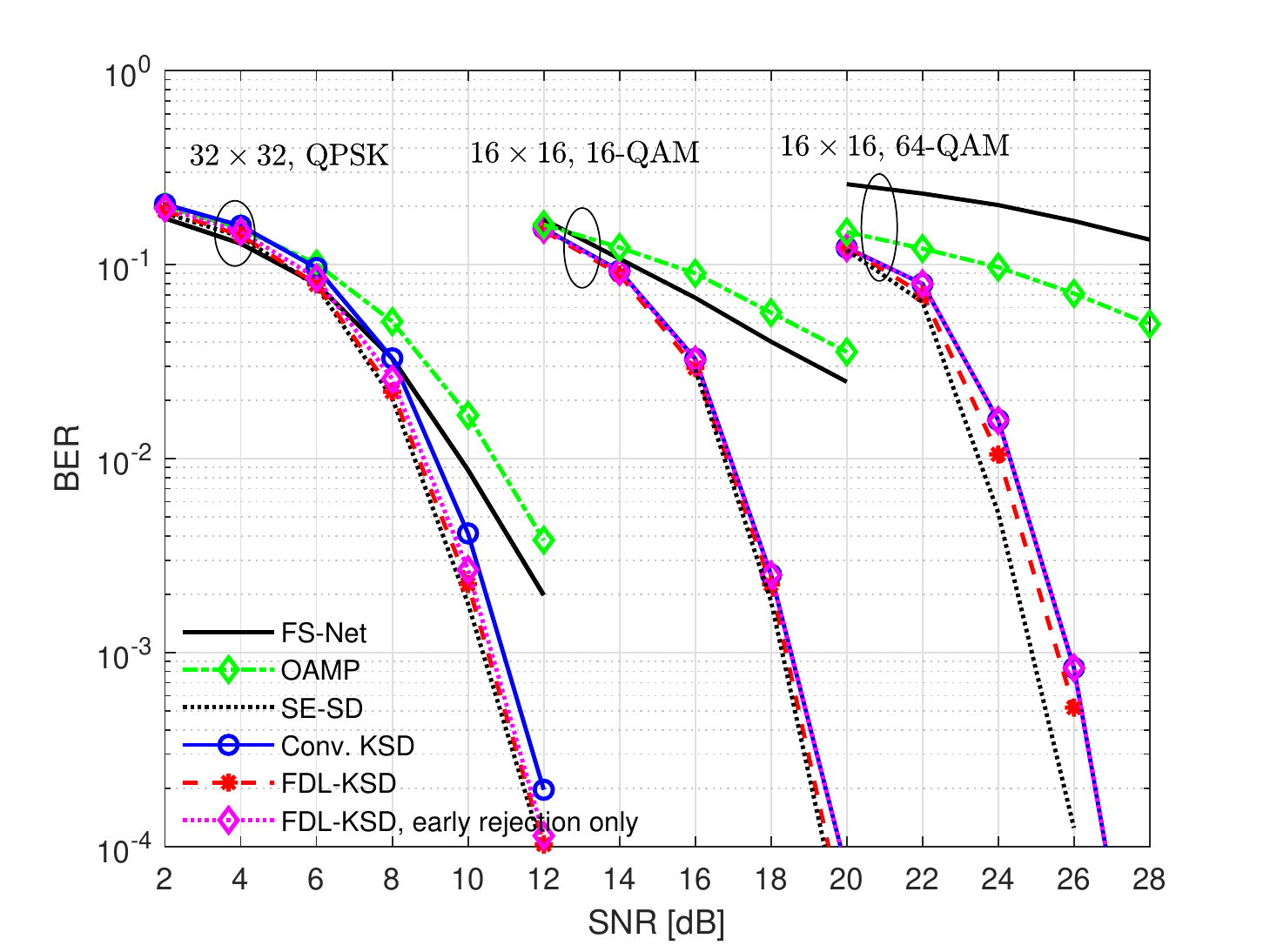}
		\caption{BER performance of the proposed FDL-KSD scheme compared to those of the conventional KSD {and OAMP schemes} for a $(32 \times 32)$ MIMO system with $L = 15$ and QPSK, a $(16 \times 16)$ MIMO system with $L = 30$ and 16-QAM, and a {$(16 \times 16)$ MIMO system with $L = 15$ and 64-QAM}, and $K=256$.}
		\label{fig_ber_KSD}
	\end{figure}
	
	\begin{figure*}[t]
		\centering
		\includegraphics[scale=0.6]{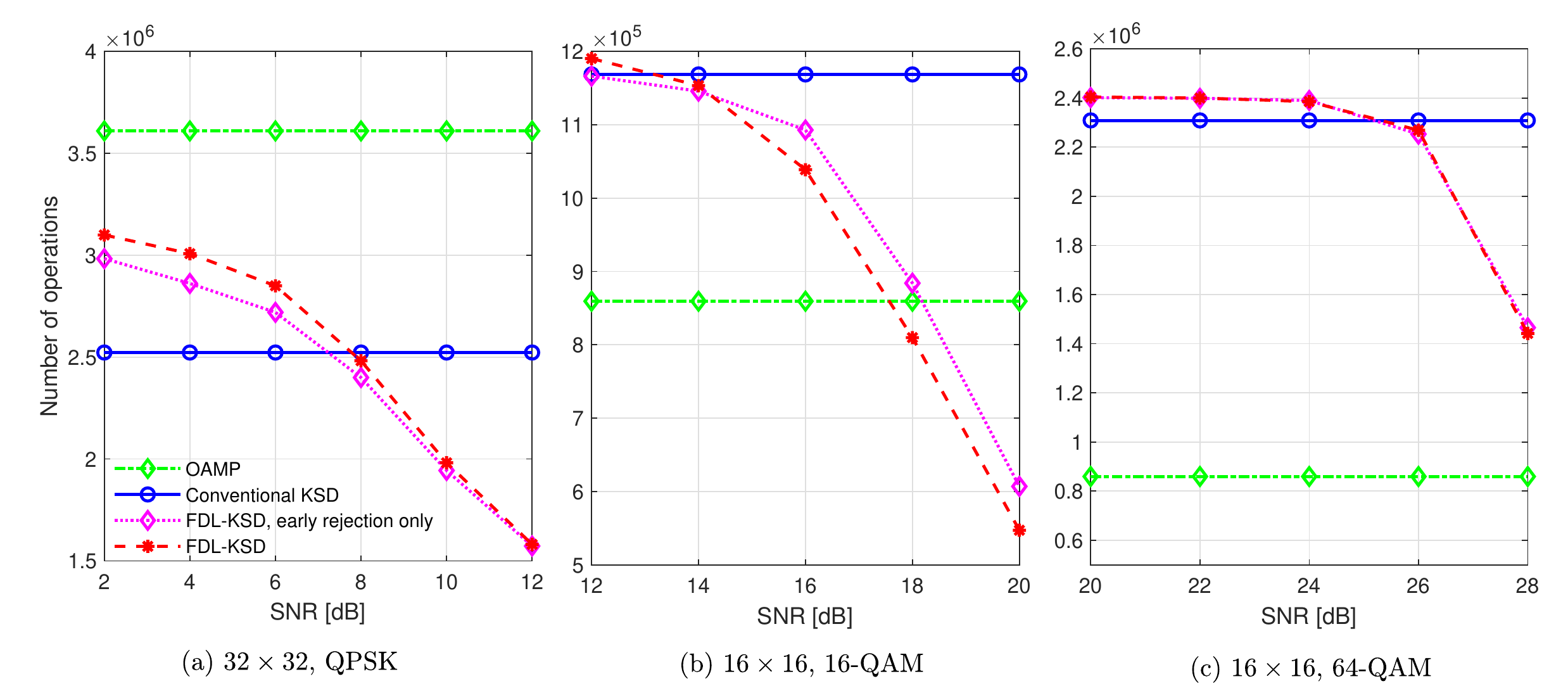}
		\caption{Computational complexity of the proposed FDL-KSD scheme compared to those of the conventional KSD {and the OAMP schemes} for a $(32 \times 32)$ MIMO system with $L = 15$ and QPSK, a $(16 \times 16)$ MIMO system with $L = 30$ and 16-QAM, and a {$(16 \times 16)$ MIMO system with $L = 15$ and 64-QAM}, and $K=256$.}
		\label{fig_comp_KSD}
	\end{figure*}
	
	In Fig. \ref{fig_ber_KSD}, we show the BER performance and complexity of the proposed FDL-KSD and the conventional KSD schemes for a $32 \times 32$ MIMO system with QPSK, a $16 \times 16$ MIMO system with 16-QAM, and a {$(16 \times 16)$ MIMO system with 64-QAM}. We note that no existing work in the literature considers the application of DL to KSD. Therefore, we only compare the performance and complexity of the proposed FDL-KSD to those of the conventional KSD {and OAMP schemes}. Furthermore, we also show the performance and complexity of the proposed FDL-KSD with early rejection only, to demonstrate that this early rejection scheme attains not only performance improvement, but also complexity reduction.
	
	\begin{figure}[t]
		\centering
		\subfigure[$32 \times 32$ MIMO, $L = 15$, SNR $=12$ dB with QPSK]
		{
			\includegraphics[scale=0.52]{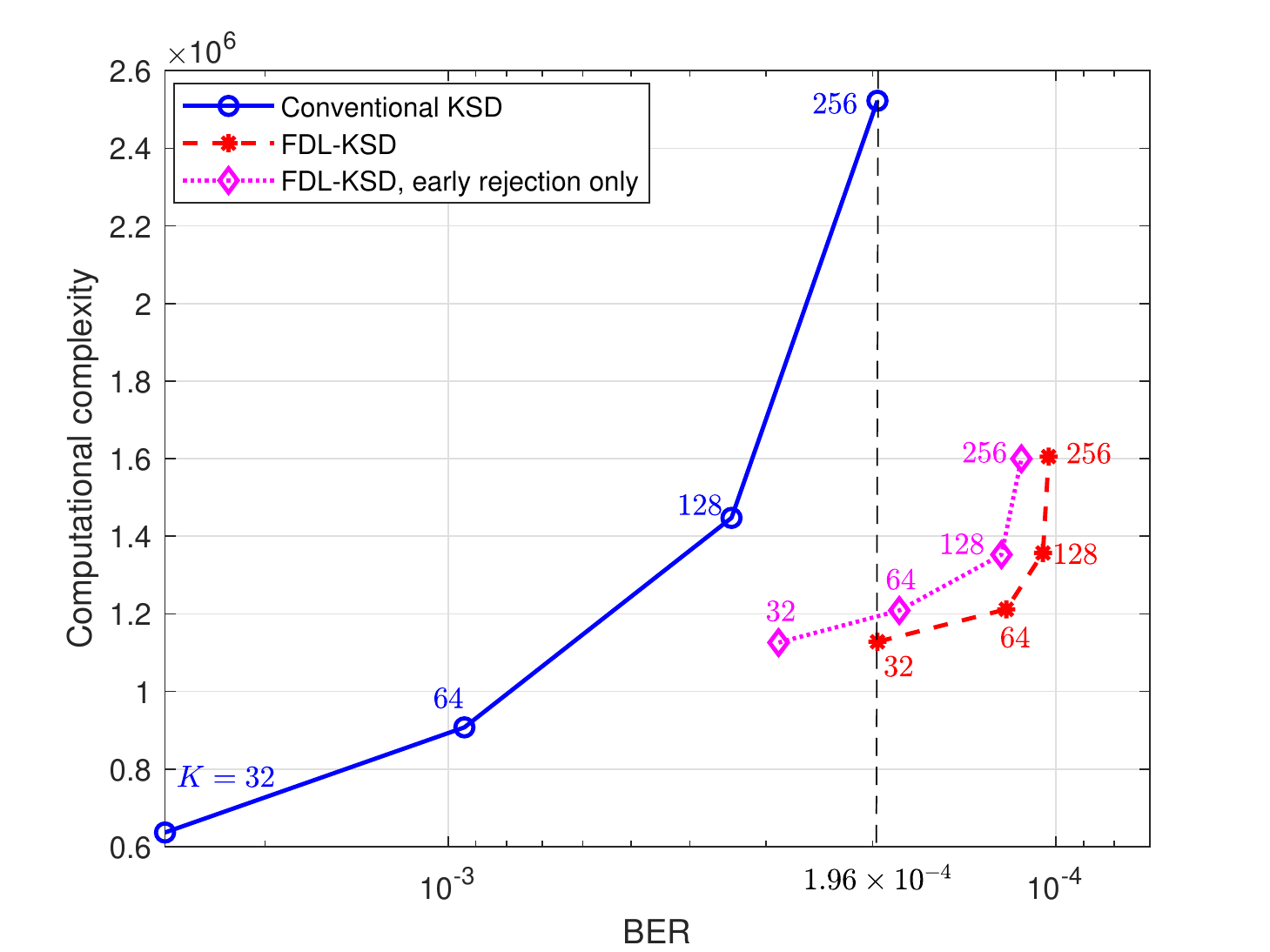}
			\label{tradeoff_KSD_QPSK}
		}
		\subfigure[$16 \times 16$ MIMO, $L = 30$, SNR $=20$ dB with 16-QAM]
		{
			\includegraphics[scale=0.52]{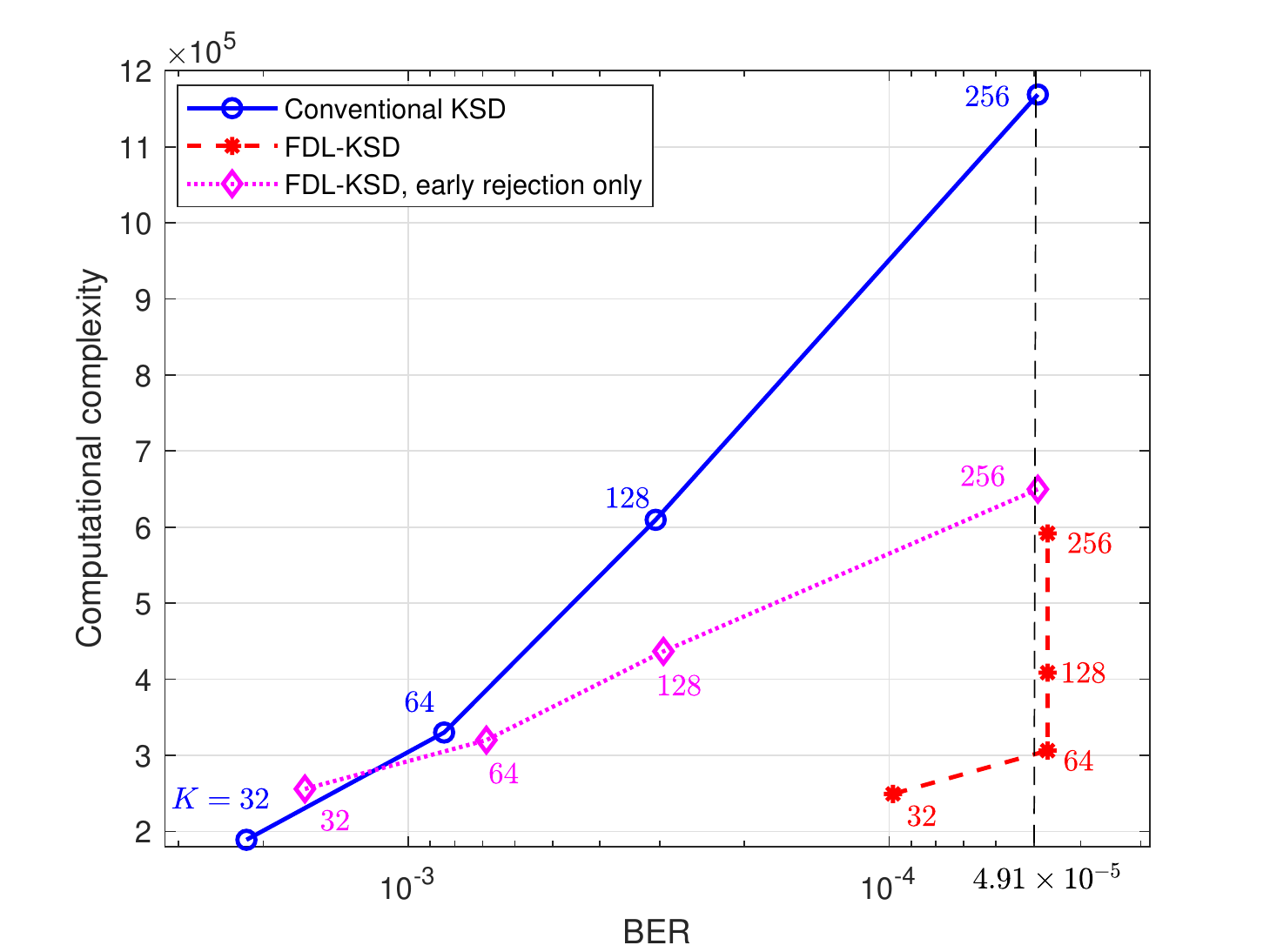}
			\label{tradeoff_KSD_QAM}
		}
		\caption{Tradeoff between BER performance and computational complexity of the proposed FDL-KSD scheme compared to that of the conventional KSD for $K=\{32, 64, 128, 256\}$.}
		\label{tradeoff_KSD}
	\end{figure}
	
	In Fig. \ref{fig_comp_KSD}, it is observed that unlike the conventional KSD scheme, whose complexity is fixed with SNRs, the proposed FDL-KSD scheme has the complexity decreasing significantly with SNRs. Specifically, the complexities of the FDL-KSD scheme in both considered systems are reduced by approximately half as the SNR increases from low to high. Moreover, it is clear that at moderate and high SNRs, the proposed FDL-KSD scheme achieves better performance with considerably lower complexity than the conventional KSD scheme. In particular, the early rejection can achieve improved performance and reduced complexity w.r.t. the conventional KSD. The additional application of candidate ordering results in further performance improvement of the FDL-KSD algorithm, as seen in Fig. \ref{fig_ber_KSD}. Specifically, in both considered systems, a performance improvement of $0.5$ dB in SNR is achieved. At the same time,  complexity reductions of $38.3\%$, $50.2\%$, {and $40\%$ w.r.t. the conventional KSD are attained at SNR $=\{12, 20, 28\}$ dB in Figs. \ref{fig_comp_KSD}(a), \ref{fig_comp_KSD}(b), and \ref{fig_comp_KSD}(c), respectively.} We note that the proposed FDL-KSD has higher or comparable complexity w.r.t. the conventional KSD at low SNRs because the complexity required for the FS-Net is included. {In particular, in Fig. \ref{fig_comp_KSD}(c), the OAMP scheme has a relatively low complexity. However, its performance is not near-optimal, whereas that of the SD-and KSD-based alternatives is.}

	\begin{figure}[t]
		\centering
		\subfigure[BER performance]
		{
			\includegraphics[scale=0.55]{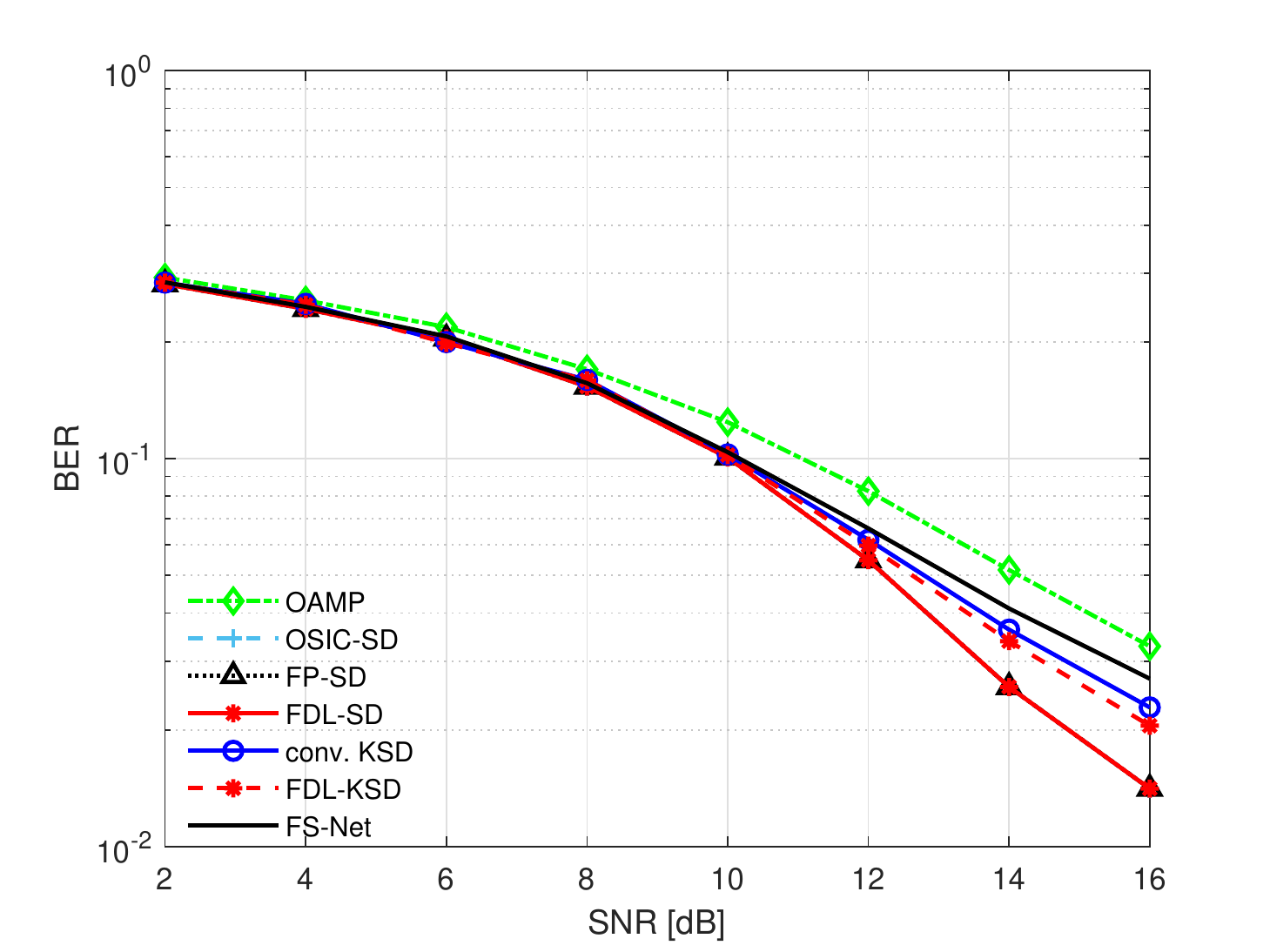}
			\label{fig_ber_onering}
		}
		\subfigure[Computational complexity]
		{
			\includegraphics[scale=0.55]{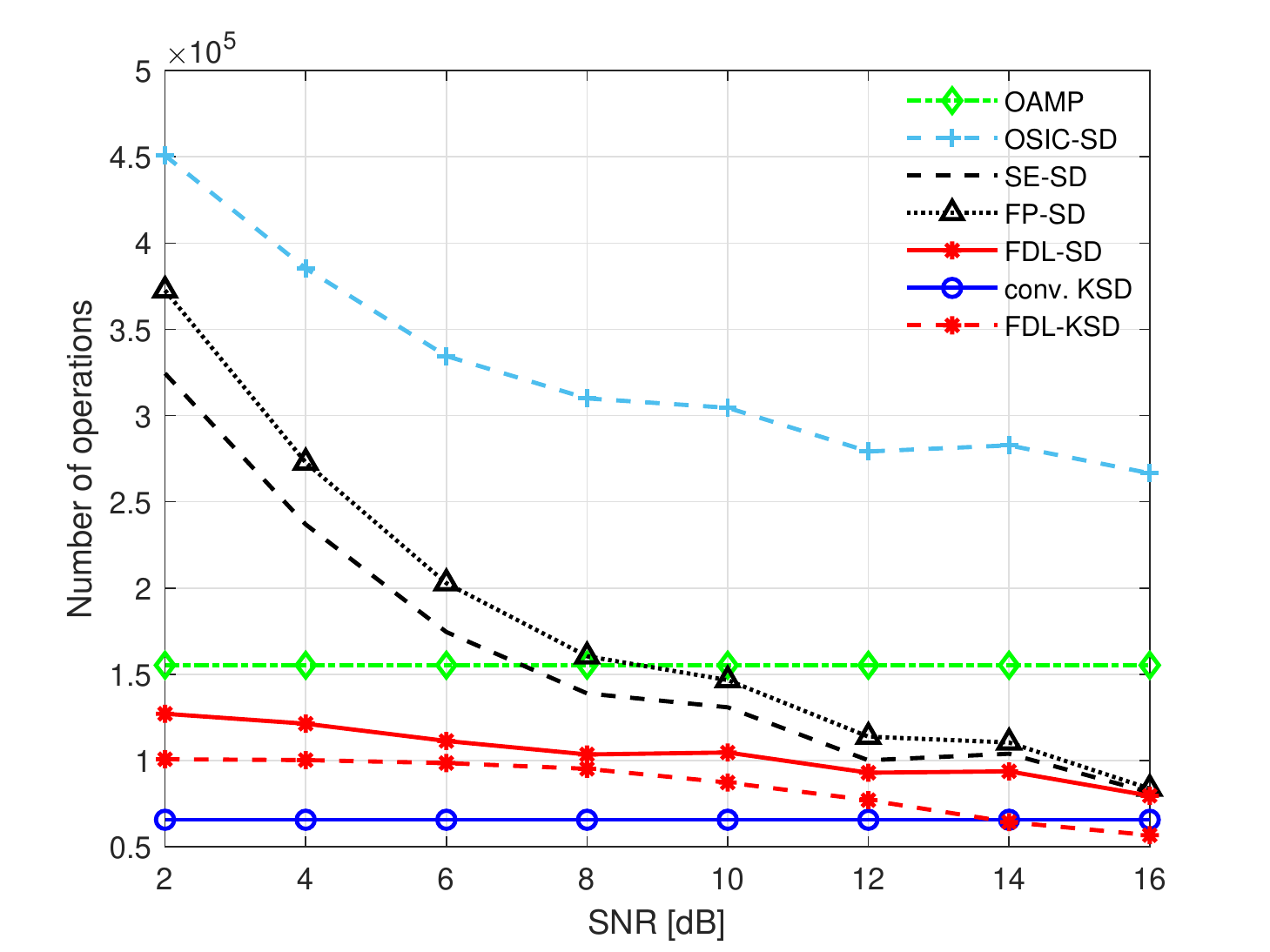}
			\label{fig_comp_onering}
		}
		\caption{{BER performance and complexity of the proposed schemes in one-ring correlated channels.}}
		\label{fig_onering}
	\end{figure}
	
	In Fig. \ref{tradeoff_KSD}, we show the improvement in the performance--complexity tradeoff of the proposed FDL-KSD scheme w.r.t. the conventional KSD scheme for a $32 \times 32$ MIMO system with QPSK and SNR $=12$ dB and a $16 \times 16$ MIMO system with 16-QAM and SNR $=20$ dB. Various values for $K$ are considered, including $K=\{32, 64, 128, 256\}$. We make the following observations:
	\begin{itemize}
		\item First, it is clear that the proposed FDL-KSD scheme not only achieves better BER performance but also requires much lower complexity than the conventional KSD scheme. For example, to attain a BER of $1.96 \times 10^{-4}$ in the $32 \times 32$ MIMO system with QPSK, the conventional KSD scheme requires $K=256$, whereas only $K<64$ is sufficient for the proposed FDL-KSD scheme with early rejection only, corresponding to a complexity reduction ratio of $52.5 \%$, and only $K=32$ is required for the FDL-KSD scheme with both early rejection and candidate ordering, resulting in $55.6 \%$ complexity reduction.
		\item Second, the complexity reduction is more significant as $K$ increases. This is because in the proposed FDL-KSD scheme, the number of actual survival nodes is not $K$, but $K_m \leq K, \forall m$, as discussed in Section V-B.
		\item Moreover, the performance--complexity tradeoff of the conventional KSD scheme significantly depends on $K$, as discussed in Section V-B. In Fig. \ref{tradeoff_KSD}, its BER performance can be improved dramatically as $K$ increases, which, however, causes considerably high complexity. In contrast, the performance--complexity tradeoff of the proposed FDL-KSD scheme is comparatively stable with $K$. For example, its BER performance in the $32 \times 32$ MIMO system with QPSK is approximately the same for $K=128$ and $K=256$, and its complexity increases relatively slowly as $K$ increases. In contrast, in the $16 \times 16$ MIMO system with 16-QAM, $K=64$ is sufficient to achieve a BER of $4.68 \times 10^{-5}$, and a further increase of $K$ to $\{128, 256\}$ does not result in performance improvement. In this case, we can conclude that $K=64$ is optimal for the FDL-KSD scheme.
	\end{itemize}

	\subsection{{Performance and Complexity of FDL-SD/KSD for Highly Correlated Channels}}

	{In Fig. \ref{fig_onering}, we show the BER performance and computational complexity of the proposed FDL-SD/KSD compared to those of the conventional FP-SD, SE-SD, OSIC-SD, KSD, FS-Net, and OAMP schemes under a highly correlated channel. Specifically, a $10 \times 16$ MIMO system with QPSK and the one-ring channel model \cite{he2020model} is assumed, which is also used to generate the training data of the FS-Net. Furthermore, we assume $L=10$, $T=2$, and $K=32$ for the number of layers in FS-Net, number of iterations in OAMP, and number of surviving nodes in KSD, respectively. It is observed that the performance of the considered schemes degrades  significantly w.r.t. the case of i.i.d. Rayleigh channels. However, it is worth noting that in highly correlated channels, the FS-Net performs similarly to the KSD/SD, in contrast with the DetNet, OAMP-Net, and LcgNet in \cite{samuel2019learning, he2020model, wei2020learned}. As a result, the FDL-SD and FDL-KSD still achieve complexity reduction without any performance loss, similar to the observations for the i.i.d. Rayleigh channel, which further justifies that the FS-Net is a highly efficient scheme to generate initial solutions for the proposed FDL-SD/KSD schemes.}

	\section{Conclusion}
	In this paper, we have presented a novel application of DL to both the conventional SD and KSD, resulting in the FDL-SD and FDL-KSD schemes, respectively. The main idea is to leverage the FS-Net to generate a highly reliable initial solution with low complexity. The initial solution determined by the FS-Net is exploited for candidate and layer ordering in the FDL-SD scheme, and for early rejection and layer ordering in the FDL-KSD scheme. Unlike the existing DL-aided SD schemes, the proposed application of DL to the SD schemes does not require performing the conventional SD schemes to generate the training data. Therefore, the employed DNN, i.e., FS-Net, can be trained with significantly less time and computational resources than those required in existing works. Our simulation results justify the performance and complexity-reduction gains of the proposed schemes. Specifically, the FDL-SD scheme achieves remarkable complexity reduction, which exceeds $90\%$, without any performance loss. Moreover, the proposed FDL-KSD scheme attains a dramatically improved performance--complexity tradeoff. {We note that the proposed applications of DL to SD/KSD are not limited to the use of the FS-Net, and a scheme with a superior performance--complexity tradeoff can potentially provide higher complexity reduction gains, which motivates further developments of improved DL models for signal detection.}
	
	\bibliographystyle{IEEEtran}
	\bibliography{IEEEabrv,Bibliography}
	
\end{document}